%% file: SciPostPhys.tex
\documentclass{SciPost}
\input{utils/preamble_scipost}
\input{utils/macros_CMS_TDR.tex}

\input{utils/macros_user.tex}
\hypersetup{
    colorlinks,
    linkcolor={red!50!black},
    citecolor={blue!50!black},
    urlcolor={blue!80!black}
}

\newcommand{\magenta}[1]{\color{magenta} #1 \color{black}}

\usepackage[bitstream-charter]{mathdesign}
\usepackage{url}
\usepackage{color}
\usepackage{fontawesome}
\usepackage{booktabs}

\DeclareSymbolFont{usualmathcal}{OMS}{cmsy}{m}{n}
\DeclareSymbolFontAlphabet{\mathcal}{usualmathcal}

\fancypagestyle{SPstyle}{
\fancyhf{}
\lhead{\colorbox{scipostblue}{\bf \color{white} ~SciPost Physics Community Reports }}
\rhead{{\bf \color{scipostdeepblue} ~Submission }}

\fancyfoot[C]{\textbf{\thepage}}
}

\begin{document}\vspace*{-1cm}
\begin{flushleft} 
\magenta{LHCHWG-2026-007\\
TIF-UNIMI-2025-28}
\end{flushleft}

\pagestyle{SPstyle}

\newcommand{\CB}[1]{{\textcolor{brown}{{\it CB: #1}}}}
\newcommand{\TB}[1]{{\textcolor{red}{{\it TB: #1}}}}

\input{CBdefinitions}

\begin{center}{\Large \textbf{\color{scipostdeepblue}{
Simulations and flavour-scheme studies for Higgs-boson production in association with charm quarks\\
}}}\end{center}

\begin{center}\textbf{
Tiziano Bevilacqua\textsuperscript{1$\star$},
Christian Biello\textsuperscript{2$\S$}, Lea Michela Caminada\textsuperscript{3,4$\times$},\\
Clemens Lange\textsuperscript{4$\triangle$}, Marino Missiroli\textsuperscript{5$\circ$}, 
Davide Pagani\textsuperscript{6$\dagger$},
Michele Selvaggi\textsuperscript{7$\ddagger$}
and~Marco~Zaro\textsuperscript{8$\diamond$}
}\end{center}

\begin{center}
{\bf 1} Institute for Particle Physics and Astrophysics, ETH Zürich, 8093 Zürich, Switzerland
\\
{\bf 2} Institute for Theoretical Physics, ETH Zurich, 8093 Zürich, Switzerland
\\
{\bf 3} Physik-Institut, University of Zürich, 8057 Zurich, Switzerland
\\
{\bf 4} PSI Center for Neutron and Muon Sciences, Forschungsstrasse 111, 5232 Villigen, Switzerland
\\
{\bf 5} Dipartimento di Fisica e Astronomia G. Galilei, Università di Padova, Via F Marzolo, 35131 Padova, Italy
\\
{\bf 6} INFN, Sezione di Bologna, Via Irnerio 46, Bologna, I-40126, Italy
\\
{\bf 7} CERN
\\
{\bf 8} INFN, Sezione di Milano and Università degli Studi di Milano, 
Via Celoria 16, Milano, I-20133, Italy
\\[\baselineskip]
$\star$ \href{mailto:tiziano.bevilacqua@cern.ch}{\small tiziano.bevilacqua@cern.ch}\,,\quad
$\S$ \href{mailto:cbiello@phys.ethz.ch}{\small cbiello@phys.ethz.ch}\,,\quad$\times$ \href{mailto:lea.michaela.caminada@cern.ch}{\small lea.michaela.caminada@cern.ch}\,,\quad\\
$\triangle$ \href{mailto:clemens.lange@cern.ch}{\small clemens.lange@cern.ch}\,,\quad$\circ$ \href{mailto:marino.missiroli@cern.ch}{\small marino.missiroli@cern.ch}\,,\quad$\dagger$ \href{mailto:davide.pagani@bo.infn.it}{\small davide.pagani@bo.infn.it}\,,\quad\\$\ddagger$ \href{mailto:michele.selvaggi@cern.ch}{\small michele.selvaggi@cern.ch}\,,\quad
$\diamond$ \href{mailto:marco.zaro@mi.infn.it}{\small marco.zaro@mi.infn.it}
\end{center}

\section*{\color{scipostdeepblue}{Abstract}}
\textbf{\boldmath{%
We present a detailed NLO+PS study of Higgs-boson production in association with charm and bottom quarks, with particular focus on the modelling of the \hpc final state. We systematically compare predictions in massive and massless flavour schemes, quantify scale and flavour-scheme uncertainties, and assess the impact of interference and loop-induced contributions. Special emphasis is placed on the Higgs production via charm Yukawa fusion, for which we compare NLO+PS predictions with a NNLO+PS calculation of $\text{c}\overline{\text{c}}\text{H}$ production in the four-flavour scheme at 13.6 TeV. Based on these results, we provide the first practical recommendations for the simulation of \PQc\PAQc\PH\ production in LHC analyses.
}}

\vspace{\baselineskip}

\noindent\textcolor{white!90!black}{%
\fbox{\parbox{0.975\linewidth}{%
\textcolor{white!40!black}{\begin{tabular}{lr}%
  \begin{minipage}{0.6\textwidth}%
    {\small Copyright attribution to authors. \newline
    This work is a submission to SciPost Physics. \newline
    License information to appear upon publication. \newline
    Publication information to appear upon publication.}
  \end{minipage} & \begin{minipage}{0.4\textwidth}
    {\small Received Date \newline Accepted Date \newline Published Date}%
  \end{minipage}
\end{tabular}}
}}
}


\vspace{10pt}
\noindent\rule{\textwidth}{1pt}
\tableofcontents
\noindent\rule{\textwidth}{1pt}
\vspace{10pt}

\input{sections/introduction.tex}
\input{sections/simulation.tex}

\input{sections/conclusions.tex}

\section*{Acknowledgements}

C.B. would like to thank Aparna Sankar, Marius Wiesemann and Giulia Zanderighi for previous work on \minnlo{} simulations.
T.B. acknowledges funding from the Swiss National Science Foundation under contracts number 200021-236405.
C.B. acknowledges funding from the Swiss National Science Foundation, grant 10001706.
L.C. acknowledge the financial support of the Swiss National Science Foundation Grant No~PP00P2\_150556. 
M.Z. acknowledges financial support by the MUR (Italy), with funds of the European Union (NextGenerationEU), through the PRIN2022 grant 2022EZ3S3F.
We have used the Max Planck Computing and Data Facility (MPCDF) in Garching to carry out the NNLO+PS simulations.

\bibliography{scipost/bibliography}

\appendix
\input{sections/Appendix}

\end{document}

%% file: utils/macros_CMS_TDR.tex
\newcommand{\etal}   {\mbox{et al.}\xspace} 
\newcommand{\ie}     {\mbox{i.e.}\xspace}

\newcommand{\unit}[1]{{\ensuremath{\text{\,#1}}}\xspace}

\newcommand{\GeV}    {{\ensuremath{\,\text{Ge\hspace{-.08em}V}}}\xspace}
\newcommand{\TeV}    {{\ensuremath{\,\text{Te\hspace{-.08em}V}}}\xspace}

\newcommand{\fb}  {{\mbox{\ensuremath{\,\text{fb}}}}\xspace}

\newcommand{\PYTHIA} {{\textsc{pythia}}\xspace}

\newcommand{\MADGRAPH} {\textsc{Mg5\_aMC}\xspace}

\newcommand{\MGvATNLO}{\textsc{MadGraph5\_aMC@NLO}\xspace}
\newcommand{\POWHEG} {{\textsc{powheg}}\xspace}

\usepackage{scalerel}
\newsavebox{\foobox}
\newcommand{\slantbox}[2][0]{\mbox{%
        \sbox{\foobox}{#2}%
        \hskip\wd\foobox
        \pdfsave
        \pdfsetmatrix{1 0 #1 1}%
        \llap{\usebox{\foobox}}%
        \pdfrestore
}}
\newcommand\unslant[2][-.25]{%
  \mkern1.2mu%
  \ThisStyle{\slantbox[#1]{$\SavedStyle#2$}}%
  \mkern-1.2mu%
}

\newcommand{\uptau}{\unslant\tau}

\newcommand{\Pe}{{\ensuremath{\text{e}}}\xspace} 
\newcommand{\PGt}{{\ensuremath{\uptau}}\xspace}

\newcommand{\PQu}{{\ensuremath{\text{u}}}\xspace} 
\newcommand{\PQd}{{\ensuremath{\text{d}}}\xspace} 
\newcommand{\PQc}{{\ensuremath{\text{c}}}\xspace} 
\newcommand{\PQs}{{\ensuremath{\text{s}}}\xspace} 
\newcommand{\PQt}{{\ensuremath{\text{t}}}\xspace} 
\newcommand{\PQb}{{\ensuremath{\text{b}}}\xspace} 
\newcommand{\PAQc}{{\ensuremath{\bar{\text{c}}}}\xspace} 
\newcommand{\PAQt}{{\ensuremath{\bar{\text{t}}}}\xspace} 
\newcommand{\PAQb}{{\ensuremath{\bar{\text{b}}}}\xspace} 
\newcommand{\PQq}{{\ensuremath{q}}\xspace} 
\newcommand{\Pg}{{\ensuremath{g}}\xspace} 
\newcommand{\PW}{{\ensuremath{\text{W}}}\xspace} 
\newcommand{\PZ}{{\ensuremath{\text{Z}}}\xspace} 
\newcommand{\PH}{{\ensuremath{\text{H}}}\xspace} 
\newcommand{\Zg}{{\ensuremath{\PZ\kern-0.5pt/\kern-0.5pt\gamma^*}}\xspace} 
\newcommand{\Pp}{\ensuremath{\mathrm{p}}\xspace} 

\newcommand{\kt}     {\ensuremath{k_\text{T}}\xspace}
\newcommand{\pt}     {\ensuremath{p_\text{T}}\xspace}

\newcommand{\ptmiss} {\ensuremath{\pt^\text{miss}}\xspace}
\newcommand{\ptvecmiss}{\ensuremath{{\myvec{p}}_{\mathrm{T}}^{\kern1pt\text{miss}}}\xspace}
\newcommand{\ptvecvis}[1][]{\ensuremath{{\myvec{p}}_{\mathrm{T}\ifthenelse{\isempty{#1}}{}{,#1}}^{\kern1pt\text{vis}}}\xspace}

\newcommand{\mT}     {{\ensuremath{m_\text{T}}}\xspace}

\newcommand{\ch} {\ensuremath{\PQc\PH}\xspace}
\newcommand{\cch} {\ensuremath{\PQc\PAQc\PH}\xspace}
\newcommand{\bbh} {\ensuremath{\PQb\PAQb\PH}\xspace}
\newcommand{\bh} {\ensuremath{\PQb\PH}\xspace}

\newcommand{\hpc} {\ensuremath{\PH+\PQc}\xspace}
\newcommand{\hpb} {\ensuremath{\PH+\PQb}\xspace}

\newcommand{\kappac}  {\ensuremath{\kappa_c}\xspace}

\newcommand{\kappab}  {\ensuremath{\kappa_b}\xspace}
\newcommand{\kappai}  {\ensuremath{\kappa_i}\xspace}

%% file: utils/macros_user.tex
\newcommand*{\myvec}[1]{\vec{\mkern0mu#1}} 

\newcommand{\tauh}   {{\ensuremath{\PGt\kern-0.5pt_\text{h}}}\xspace} 
\providecommand{\tt} {} 
\renewcommand{\tt}   {{\ensuremath{\PGt\PGt}\xspace}}
\newcommand{\ee}     {{\ensuremath{\Pe\Pe}}\xspace}

\newcommand{\tautau} {{\ensuremath{\PGt\kern-0.2pt\PGt}}\xspace}
\newcommand{\ditau}  {{\ensuremath{\tauh\kern-0.5pt\tauh}}\xspace}

\newcommand{\muR}    {{\ensuremath{\mu_\text{R}}\xspace}} 
\newcommand{\muF}    {{\ensuremath{\mu_\text{F}}\xspace}} 
\newcommand{\mll}    {{\ensuremath{m_{\ell\ell}}}\xspace}



%% file: CBdefinitions.tex
\newcommand{\MSbar}{\ensuremath{\overline{\text{MS}}}}
\newcommand{\fonll}{FONLL}
\newcommand{\nlonnllpart}{NLO+NNLL\textsubscript{part}+$y_by_t$}
\newcommand{\nnnres}{$\text{N}^3\text{LL}'+\text{aN}^3\text{LO}$}

\providecommand{\href}[2]{#2}

\newcommand{\incl}{{\tt inclusive}}
\newcommand{\fidYR}{{\tt fiducial-YR}}
\newcommand{\fidATLAS}{{\tt fiducial-ATLAS}}

\newcommand{\stepone}{{Step\,I}}
\newcommand{\steptwo}{{Step\,II}}
\newcommand{\stepthree}{{Step\,III}}

\newcommand\tS{\tilde{S}}
\newcommand\F{${\textrm F}$}
\newcommand\FJ{${\textrm FJ}$}
\newcommand\FJJ{${\textrm FJJ}$}
\newcommand\PhiBorn{\Phi_{\scriptscriptstyle \textrm B}}
\newcommand\PhiReal{\Phi_{\scriptscriptstyle \textrm R}}
\newcommand\PhiB{\Phi_{\scriptscriptstyle \textrm F}}
\newcommand\PhiBres{\Phi_{\scriptscriptstyle \textrm F,res}}

\newcommand{\flav}{\ell}
\newcommand{\flavBorn}{\flav_{\scriptscriptstyle \textrm B}}
\newcommand{\fullflavBorn}{\hat \flav_{\scriptscriptstyle \textrm B}}
\newcommand{\flavprimeBorn}{\flav'_{\scriptscriptstyle \textrm B}}
\newcommand{\fullflavprimeBorn}{\hat \flav'_{\scriptscriptstyle \textrm B}}
\newcommand{\flavB}{\flav_{\scriptscriptstyle \textrm F}}
\newcommand{\fullflavB}{\hat \flav_{\scriptscriptstyle \textrm F}}
\newcommand{\flavBJ}{\flav_{\scriptscriptstyle \textrm FJ}}
\newcommand{\fullflavBJ}{\hat \flav_{\scriptscriptstyle \textrm FJ}}
\newcommand{\flavBJJ}{\flav_{\scriptscriptstyle \textrm FJJ}}
\newcommand{\fullflavBJJ}{\hat \flav_{\scriptscriptstyle \textrm FJJ}}
\newcommand{\projflav}{\flavB\leftarrow\flavBJ}
\newcommand{\CF}{C_{\mathrm{F}}}
\newcommand{\CA}{C_{\mathrm{A}}}
\newcommand{\NC}{N_{\mathrm{c}}}
\newcommand{\nf}{N_f}
\newcommand{\TF}{T_{\mathrm{F}}}

\newcommand{\flavZg}{\flav_{\scriptscriptstyle Z\gamma}}
\newcommand{\fullflavZg}{\hat \flav_{\scriptscriptstyle Z\gamma}}
\newcommand{\flavZgJ}{\flav_{\scriptscriptstyle Z\gamma {\textrm J}}}
\newcommand{\fullflavZgJ}{\hat \flav_{\scriptscriptstyle Z\gamma {\textrm J}}}
\newcommand{\flavZgJJ}{\flav_{\scriptscriptstyle Z\gamma {\textrm J}}}
\newcommand{\fullflavZgJJ}{\hat \flav_{\scriptscriptstyle Z\gamma {\textrm J}}}
\newcommand{\projflavZg}{\flavZg\leftarrow\flavZgJ}

\newcommand\PhiBJ{\Phi_{\scriptscriptstyle \textrm FJ}}
\newcommand\ZJ{Z\gamma J}
\newcommand\PhiZJ{\Phi_{\scriptscriptstyle \textrm Z\gamma J}}
\newcommand\PhiBJbar{{\bar \Phi}'_{\scriptscriptstyle \textrm FJ}}
\newcommand\PhiBJJ{\Phi_{\scriptscriptstyle \textrm FJJ}}
\newcommand\PhiZJJ{\Phi_{\scriptscriptstyle \textrm Z\gamma JJ}}
\newcommand\PhiZgam{\Phi_{\scriptscriptstyle \textrm Z\gamma}}
\newcommand{\Fcorr}{F^{\tmop{corr}}_\ell}
\renewcommand{\order}[1]{{\cal O}\left(#1\right)}
\newcommand{\aew}{\alpha_{\text{\scalefont{0.77}EW}}} 
\newcommand{\aw}{\alpha_w}
\newcommand{\asCMW}{\alpha_s^{\textrm †CMW}}
\newcommand{\NNLL}{\text{NNLL}}
\newcommand{\eff}{\epsilon}
\providecommand{\ee}{\ell^+\ell^-}
\providecommand{\kt}[1]{k_{\scaleto{\textrm T}{4pt},#1}}
\newcommand{\veckt}[1]{\vec{k}_{\scaleto{\textrm T}{4pt},#1}}
\newcommand{\fullF}{\mathcal{F}}
\newcommand{\FNLL}{\mathcal{F}_{\textrm NLL}}
\newcommand{\FNNLL}{\mathcal{F}_{\textrm NNLL}}
\providecommand{\ie}{i.e.\,}
\newcommand{\css}{\text{\scriptsize CSS}}
\newcommand{\zi}{z_i^{(\ell_i)}}
\providecommand{\pt}{p_{\text{\relscale{0.77}T}}}
\newcommand{\GZ}{{\Gamma_Z}}
\newcommand{\GW}{{\Gamma_W}}
\newcommand{\thW}{{\theta_W}}
\newcommand{\mtop}{{m_{\text{\relscale{0.77}top}}}}
\newcommand{\qt}{{q_{\text{\relscale{0.77}T}}}}
\newcommand{\ptarg}[1]{{p_{\text{\relscale{0.77}T,$#1$}}}}
\newcommand{\marg}[1]{{m_{\text{\relscale{0.77}$#1$}}}}
\newcommand{\ptg}{p_{\text{\relscale{0.77}T,$\gamma$}}}
\newcommand{\ptgcut}{{p_{\text{\relscale{0.77}T,$\gamma$}}^{\textrm cut}}}
\newcommand{\ptjcut}{{p_{\text{\relscale{0.77}T,$j$}}^{\textrm cut}}}

\newcommand{\ptjmin}{\bar{p}_{\text{\relscale{0.77}T,$j$}}}
\newcommand{\ptgmin}{\bar{p}_{\text{\relscale{0.77}T,$\gamma$}}}
\newcommand{\ptgone}{p_{\text{\relscale{0.77}T,$\gamma_1$}}}
\newcommand{\ptgtwo}{p_{\text{\relscale{0.77}T,$\gamma_2$}}}
\newcommand{\ptgthree}{p_{\text{\relscale{0.77}T,$\gamma_3$}}}
\newcommand{\ptH}{p_{\text{\relscale{0.77}T,$H$}}}
\newcommand{\ptHjj}{p_{\text{\relscale{0.77}T,$Hjj$}}}
\newcommand{\mjj}{m_{\text{\relscale{0.77}$jj$}}}

\newcommand{\ptr}{{p_{\text{\relscale{0.77}T}}}^{\text{\relscale{0.9}r}}}
\newcommand{\ptrad}{{p_{\text{\relscale{0.77}T,rad}}}}
\newcommand{\pth}{{p_{\text{\relscale{0.77}T,$H$}}}}
\newcommand{\ptz}{{p_{\text{\relscale{0.77}T,$Z$}}}}
\newcommand{\ptw}{{p_{\text{\relscale{0.77}T,$W$}}}}
\newcommand{\ptnu}{{p_{\text{\relscale{0.77}T,$\nu$}}}}
\newcommand{\mtwz}{{m_{\text{\relscale{0.77}T,$WZ$}}}}
\newcommand{\dphiwz}{{\Delta\phi_{\text{\relscale{0.77}$WZ$}}}}
\newcommand{\dyZlW}{{|y_{\text{\relscale{0.77}$Z$}}-y_{\text{\relscale{0.77}$
\ell_W$}}|}}
\newcommand{\ptww}{{p_{\text{\relscale{0.77}T,$WW$}}}}
\newcommand{\ptwp}{{p_{\text{\relscale{0.77}T,$W^+$}}}}
\newcommand{\ptwm}{{p_{\text{\relscale{0.77}T,$W^-$}}}}
\newcommand{\mtww}{{m_{\text{\relscale{0.77}T,$WW$}}}}
\newcommand{\mtwwexp}{{m_{\text{\relscale{0.77}T,$WW$}}^{\textrm exp}}}
\newcommand{\ptllg}{{p_{\text{\relscale{0.77}T,}\ell\ell\gamma}}}
\newcommand{\ptzg}{\ptllg}
\newcommand{\ptll}{{p_{\text{\relscale{0.77}T,$e^+e^-$}}}}
\newcommand{\ptlnu}{{p_{\text{\relscale{0.77}T,$\mu\nu_\mu$}}}}
\newcommand{\ptj}{p_{\text{\relscale{0.77}T,$j$}}}
\newcommand{\ptjone}{{p_{\text{\relscale{0.77}T,$j_1$}}}}
\newcommand{\ptjoneveto}{{p_{\text{\relscale{0.77}T,$j_1$}}^{\textrm veto}}}
\newcommand{\ptjtwo}{{p_{\text{\relscale{0.77}T,$j_2$}}}}
\newcommand{\ptlm}{{p_{\text{\relscale{0.77}T,$\ell^-$}}}}
\newcommand{\ptl}{{p_{\text{\relscale{0.77}T,$\ell$}}}}
\newcommand{\ptlone}{{p_{\text{\relscale{0.77}T,$\ell_1$}}}}
\newcommand{\ptltwo}{{p_{\text{\relscale{0.77}T,$\ell_2$}}}}
\providecommand{\ptmiss}{{p_{\text{\relscale{0.77}T,miss}}}}
\newcommand{\ptmissrel}{{p_{\text{\relscale{0.77}T,miss,rel}}}}
\newcommand{\yh}{{y_{\text{\relscale{0.77}H}}}}
\newcommand{\yz}{{y_{\text{\relscale{0.77}Z}}}}
\newcommand{\ywp}{{y_{\text{\relscale{0.77}$W^+$}}}}
\newcommand{\yl}{{y_{\text{\relscale{0.77}$\ell$}}}}
\newcommand{\ylone}{{y_{\text{\relscale{0.77}$\ell_1$}}}}
\newcommand{\dphill}{{\Delta\phi_{\text{\relscale{0.77}$\ell_1\ell_2$}}}}
\newcommand{\yww}{{y_{\text{\relscale{0.77}$WW$}}}}
\newcommand{\yjone}{{y_{\text{\relscale{0.77}$j_1$}}}}
\newcommand{\dyww}{{\Delta y_{\text{\relscale{0.77}$W^-,W^+$}}}}
\newcommand{\mh}{{m_{\text{\relscale{0.77}H}}}}
\newcommand{\mz}{{m_{\text{\relscale{0.77}Z}}}}
\newcommand{\mw}{{m_{\text{\relscale{0.77}W}}}}
\newcommand{\mww}{{m_{\text{\relscale{0.77}WW}}}}
\newcommand{\mwsq}{{m^2_{\text{\relscale{0.77}W}}}}
\newcommand{\mt}{{m_{\text{\relscale{0.77}t}}}}
\providecommand{\mT}{{m_{\text{\relscale{0.77}T}}}}
\newcommand{\mgj}{{m_{\text{\relscale{0.77}$\gamma j_1$}}}}
\newcommand{\mllg}{{m_{\text{\relscale{0.77}$\ell\ell\gamma$}}}}
\newcommand{\mlnu}{{m_{\text{\relscale{0.77}$\mu\nu_\mu$}}}}
\providecommand{\etal}{{\eta_{\text{\relscale{0.77}$\ell$}}}}
\newcommand{\etallg}{{\eta_{\text{\relscale{0.77}$\ell\ell\gamma$}}}}
\newcommand{\etalone}{{\eta_{\text{\relscale{0.77}$\ell_1$}}}}
\newcommand{\etaltwo}{{\eta_{\text{\relscale{0.77}$\ell_2$}}}}
\newcommand{\etag}{{\eta_{\text{\relscale{0.77}$\gamma$}}}}
\newcommand{\etaj}{{\eta_{\text{\relscale{0.77}j}}}}
\newcommand{\etah}{{\eta_{\text{\relscale{0.77}H}}}}
\newcommand{\detallgj}{{\Delta\eta_{\text{\relscale{0.77}$\ell\ell\gamma, j_1$}}}}
\newcommand{\dphillg}{{\Delta\phi_{\text{\relscale{0.77}$\ell\ell,\gamma$}}}}
\newcommand{\drlg}{{\Delta R_{\text{\relscale{0.77}$\ell \gamma$}}}}
\newcommand{\drgjone}{{\Delta R_{\text{\relscale{0.77}$\gamma j_1$}}}}
\newcommand{\drgjtwo}{{\Delta R_{\text{\relscale{0.77}$\gamma j_2$}}}}
\newcommand{\drej}{{\Delta R_{\text{\relscale{0.77}$e j$}}}}

\newcommand{\rjg}{R_{\text{\relscale{0.77}$j \gamma$}}}
\newcommand{\rjgone}{R_{\text{\relscale{0.77}$j \gamma_1$}}}
\newcommand{\rjgtwo}{R_{\text{\relscale{0.77}$j \gamma_2$}}}
\newcommand{\rjgthree}{R_{\text{\relscale{0.77}$j \gamma_3$}}}
\newcommand{\rjgmin}{\bar{R}_{\text{\relscale{0.77}$j \gamma$}}}
\newcommand{\rjonegone}{R_{\text{\relscale{0.77}$j_1 \gamma_1$}}}
\newcommand{\rjonegtwo}{R_{\text{\relscale{0.77}$j_1 \gamma_2$}}}
\newcommand{\rjonegthree}{R_{\text{\relscale{0.77}$j_1 \gamma_3$}}}
\newcommand{\rjtwogone}{R_{\text{\relscale{0.77}$j_2 \gamma_1$}}}
\newcommand{\rjtwogtwo}{R_{\text{\relscale{0.77}$j_2 \gamma_2$}}}
\newcommand{\rjtwogthree}{R_{\text{\relscale{0.77}$j_2 \gamma_3$}}}

\newcommand{\lw}{\ensuremath{\mu}}
\newcommand{\lpw}{\ensuremath{\ell^+_{\text{\relscale{0.77}W}}}}
\newcommand{\lmw}{\ensuremath{\ell^-_{\text{\relscale{0.77}W}}}}
\newcommand{\lpmw}{\ensuremath{\ell^{\pm}_{\text{\relscale{0.77}W}}}}

\newcommand{\lz}{\ensuremath{e}}
\newcommand{\lpz}{\ensuremath{e^+}}
\newcommand{\lmz}{\ensuremath{e^-}}
\newcommand{\lpmz}{\ensuremath{e^{\pm}_{\text{\relscale{0.77}}}}}
\newcommand{\lzlead}{\ensuremath{e_{\text{\relscale{0.77}Z,1}}}}
\newcommand{\lzsubl}{\ensuremath{e_{\text{\relscale{0.77}Z,2}}}}

\newcommand{\ptlz}{\ensuremath{p_{\text{\relscale{0.77}T,\lz}}}}
\newcommand{\ptlw}{\ensuremath{p_{\text{\relscale{0.77}T,\lw}}}}
\newcommand{\ptlzlead}{\ensuremath{p_{\text{\relscale{0.77}T,\lzlead}}}}
\newcommand{\ptlzsubl}{\ensuremath{p_{\text{\relscale{0.77}T,\lzsubl}}}}

\newcommand{\mlll}{\ensuremath{m_{\text{\relscale{0.77}$3\ell$}}}}
\newcommand{\mwz}{\ensuremath{m_{\text{\relscale{0.77}WZ}}}}
\newcommand{\mtw}{\ensuremath{m_{\text{\relscale{0.77}T,W}}}}
\newcommand{\ptwz}{\ensuremath{p_{\text{\relscale{0.77}T,WZ}}}}
\newcommand{\ptlp}{\ensuremath{p_{\text{\relscale{0.77}T,\ell'}}}}
\newcommand{\dRll}{\ensuremath{\Delta R_{\text{\relscale{0.77}$\ell\ell$}}}}
\newcommand{\dRllp}{\ensuremath{\Delta R_{\text{\relscale{0.77}$\ell\ell'$}}}}
\newcommand{\etalp}{\ensuremath{\eta_{\text{\relscale{0.77}$\ell'$}}}}

\newcommand{\qcdfull}{\ensuremath{\text{NNLO}_{\textrm QCD}^{{\textrm (QCD,QED)}_{\textrm PS}}}}
\newcommand{\qcdqcd}{\ensuremath{\text{NNLO}_{\textrm QCD}^{{\textrm (QCD)}_{\textrm PS}}}}

\newcommand{\addfull}{\ensuremath{\text{NNLO}_{\textrm QCD}^{\textrm (QCD,QED)_{\textrm PS}} + \delta{\textrm NLO}_{\textrm EW}^{\textrm (QCD,QED)_{\textrm PS}}}}
\newcommand{\addqcdfull}{\ensuremath{\text{NNLO}_{\textrm QCD}^{\textrm (QCD,QED)_{\textrm PS}} + \delta{\textrm NLO}_{\textrm EW}^{\textrm (QED)_{\textrm PS}}}}
\newcommand{\addqedfull}{\ensuremath{\text{NLO}_{\textrm EW}^{\textrm (QCD,QED)_{\textrm PS}} + \delta{\textrm NNLO}_{\textrm QCD}^{\textrm (QCD)_{\textrm PS}}}}

\newcommand{\multfull}{\ensuremath{\text{NNLO}_{\textrm QCD}^{\textrm (QCD,QED)_{\textrm PS}} \times \text{K-NLO}_{\textrm EW}^{\textrm (QCD,QED)_{\textrm PS}}}}
\newcommand{\multqcdfull}{\ensuremath{\text{NNLO}_{\textrm QCD}^{\textrm (QCD,QED)_{\textrm PS}} \times \text{K-NLO}_{\textrm EW}^{\textrm (QED)_{\textrm PS}}}}
\newcommand{\multqedfull}{\ensuremath{\text{NLO}_{\textrm EW}^{\textrm (QCD,QED)_{\textrm PS}} \times \text{K-NNLO}_{\textrm QCD}^{\textrm (QCD)_{\textrm PS}}}}

\newcommand{\multmatrix}{\ensuremath{\text{NNLO}_{\textrm QCD}^{\textrm (QCD)_{\textrm PS}} \times \text{K-NLO}_{\textrm EW}^{\text{\Matrix{}}}}}
\newcommand{\QCDpEW}{\ensuremath{ \text{NNLO}_{\textrm QCD+EW}^{\textrm (QCD, QED)_{\textrm PS}}}} 
\newcommand{\QCDtEW}{\ensuremath{ \text{NNLO}_{\textrm QCDxEW}^{\textrm (QCD, QED)_{\textrm PS}}}} 
\newcommand{\QCDtEWfo}{\ensuremath{ \text{NNLO}_{\textrm QCD}^{\textrm (QCD)_{\textrm PS}} \times \text{K-NLO}_{\textrm EW}^{\textrm (f.o.)}}}

\newcommand{\drlj}{{\Delta R_{\text{\relscale{0.77}$\ell,j$}}}}
\newcommand{\drgj}{{\Delta R_{\text{\relscale{0.77}$\gamma j$}}}}
\newcommand{\drgjo}{{\Delta R_{\text{\relscale{0.77}$\gamma j_1$}}}}
\newcommand{\drgjt}{{\Delta R_{\text{\relscale{0.77}$\gamma j_2$}}}}
\newcommand{\fcl}{{E_{\text{\relscale{0.77}T}}^{\text{\relscale{0.77}cone$0.2$}}/p_{\text{\relscale{0.77}T},\gamma}}}
\providecommand{\muF}{{\mu_{\text{\relscale{0.77}F}}}}
\providecommand{\muR}{{\mu_{\text{\relscale{0.77}R}}}}
\newcommand{\muB}{{\mu_{\text{\relscale{0.77}B}}}}
\newcommand{\muFtwo}{{\mu^2_{\text{\relscale{0.77}F}}}}
\newcommand{\muRtwo}{{\mu^2_{\text{\relscale{0.77}R}}}}
\newcommand{\muFc}{{\mu_{\text{\relscale{0.77}F},0}}}
\newcommand{\muRc}{{\mu_{\text{\relscale{0.77}R},0}}}
\newcommand{\muRy}{{\mu_{\text{\relscale{0.77}R}}^{(0),y}}}
\newcommand{\muRb}{{\mu_{\text{\relscale{0.77}R}}^{(0),\alpha}}}
\newcommand{\KF}{K_{\text{\relscale{0.77}F}}}
\newcommand{\KR}{K_{\text{\relscale{0.77}R}}}
\newcommand{\KRy}{{K^y_{\text{\relscale{0.77}R}}}}
\newcommand{\KQ}{{K_{\text{\relscale{0.77}Q}}}}
\newcommand{\Q}{{Q_{\text{\relscale{0.77}$0$}}}}
\newcommand{\Qc}{{Q_{\text{\relscale{0.77}res},0}}}

\newcommand{\noun}[1]{{\scshape #1}}
\providecommand{\MADGRAPH}{\noun{MadGraph v4}}
\newcommand{\GOSAM}{\noun{GoSam 2.0}}
\providecommand{\POWHEG}{\noun{Powheg}}
\newcommand{\SuSHi}{\noun{SuSHi}}
\newcommand{\POWHEGMiNLO}{\noun{Powheg-MiNLO}}
\newcommand{\POWHEGBOX}{\noun{Powheg-Box}}
\newcommand{\POWHEGBOXRES}{\noun{Powheg-Box-Res}}
\newcommand{\POWHEGBOXVTWO}{\noun{Powheg-Box-V2}}
\newcommand{\minlobare}{{\noun{MiNLO}}}
\newcommand{\minlosimple}{{\noun{MiNLO}}}
\newcommand{\minlo}{{\noun{MiNLO$^{\prime}$}}}
\newcommand{\minnlo}{{\noun{MiNNLO$_{\textrm{PS}}$}}}
\newcommand{\GENEVA}{\noun{Geneva}}
\newcommand{\Matrix}{{\noun{Matrix}}}
\newcommand{\OpenLoops}{{\noun{OpenLoops}}}
\providecommand{\PYTHIA}[1]{\noun{Pythia{#1}}}

\newcommand{\NNLOps}{NNLO+PS}
\newcommand{\fnnlo}{NNLO}
\newcommand{\fnnnlo}{N$^3$LO}
\newcommand{\fnlo}{NLO}
\newcommand{\NLOps}{NLO+PS}
\newcommand{\setupinclusive}{{\tt inclusive setup}}
\newcommand{\setupfiducial}{{\tt fiducial setup}}
\newcommand{\setupatlas}{{\tt ATLAS setup}}

\newcommand{\abar}{\frac{\as}{2\pi}}
\newcommand{\abarmu}[1]{\frac{\as(#1)}{2\pi}}

\newcommand{\Vsc}{V}
\newcommand{\Vwa}{V_{\textrm wa}}
\newcommand{\Vfull}{V}
\newcommand{\wzc}{V_{\textrm hc}}
\newcommand{\Vr}{V_{r}}
\newcommand{\ptB}{p_{t}^{(B)}}

\newcommand{\dZ}{d{\cal Z}[\{R', k_i\}]}
\newcommand{\RpNLL}{R'_{\mathrm{NLL}}}

\newcommand{\LambdaPWG}{\Lambda_{\textrm pwg}}

\newcommand{\yll}{{y_{\text{\relscale{0.77}\ell\ell}}}}
\providecommand{\mll}{{m_{\text{\relscale{0.77}\ell\ell}}}}
\newcommand{\mQQF}{m_{Q\bar Q{\textrm F}}}
\newcommand{\muQ}{\mu_{Q}}
\newcommand{\Mdiv}{\mathcal{M}^\textrm{IR-div}}
\newcommand{\phs}{\ensuremath{\phi^{*}_\eta}\xspace}

\newcommand{\mathd}{\mathrm{d}}
\newcommand{\tmop}[1]{\ensuremath{\operatorname{#1}}}
\newenvironment{enumeratealpha}{\begin{enumerate}[a{\textup{)}}] }{\end{enumerate}}
\newenvironment{itemizedot}{\begin{itemize} \renewcommand{\labelitemi}{$\bullet$}\renewcommand{\labelitemii}{$\bullet$}\renewcommand{\labelitemiii}{$\bullet$}\renewcommand{\labelitemiv}{$\bullet$}}{\end{itemize}}
\newcommand{\Eta}{\mathrm{H}}
\newcommand{\tmverbatim}[1]{{\ttfamily{#1}}}

\def\collr{orange}
\def\colsp{blue}
\def\colmw{purple}
\def\colsk{cyan}
\def\coldr{red}
\def\colpt{red}
\def\colgz{red}
\def\coljl{magenta}
\def\colsz{violet}
\def\colcb{brown}
\def\colas{magenta}
\def\coljm{cyan}

\newcommand{\mwcom}[1]{\textit{\textcolor{\colmw}{\{MW: #1 \}}}}
\newcommand{\gzcom}[1]{\textit{\textcolor{\colpt}{\{GZ: #1 \}}}}
\newcommand{\cbcom}[1]{\textit{\textcolor{\colcb}{\{CB: #1 \}}}}
\newcommand{\ascom}[1]{\textit{\textcolor{\colas}{\{AS: #1 \}}}}
\newcommand{\jmcom}[1]{\textit{\textcolor{\coljm}{\{JM: #1 \}}}}

\newcommand{\asdel}[1]{\textcolor{\colas}{\sout{#1}}}
\newcommand{\mwdel}[1]{\textcolor{\colmw}{\sout{#1}}}
\newcommand{\gzdel}[1]{\textcolor{\colmg}{\sout{#1}}}
\newcommand{\jmdel}[1]{\textcolor{\coljm}{\sout{#1}}}

\newcommand{\mwadd}[2]{\textcolor{\colmw}{\sout{#1}#2}}
\newcommand{\asadd}[2]{\textcolor{\colas}{\sout{#1}#2}}
\newcommand{\cbadd}[2]{\textcolor{\colcb}{\sout{#1}#2}}
\newcommand{\gzadd}[2]{\textcolor{\colgz}{\sout{#1}#2}}
\newcommand{\jmadd}[2]{\textcolor{\coljm}{\sout{#1}#2}}

\def\ltap{\raisebox{-.6ex}{\rlap{$\,\sim\,$}} \raisebox{.4ex}{$\,<\,$}} 
\def\gtap{\raisebox{-.6ex}{\rlap{$\,\sim\,$}} \raisebox{.4ex}{$\,>\,$}} 
\def\lra{\leftrightarrow} 
\def\naive{na\"{\i}ve} 
\def\bom#1{{\mbox{\boldmath $#1$}}} 
\def\to{\rightarrow}
\def\ito{\leftarrow} 
\def\nn{\nonumber} 
\def\mbbggs{m_{2b2\gamma}^{\star}}
\def\arrowlimit#1{\mathrel{\mathop{\longrightarrow}\limits_{#1}}} 
\def\ptmin{p_{T}^{\textrm min}}
\def\ptmax{p_{T}^{\textrm max}}
\def\ptveto{p_{T,\ell\ell\gamma}^{\textrm veto}}
\def\ep{\epsilon}
\def\ms{${\overline {\textrm MS}}$}
\def\perc{\%}
\def\mH{m_H}
\def\mb{m_b}
\def\mt{m_t}
\def\mw{m_W}
\def\mz{m_Z}
\def\qT{q_T}
\def\GeV{\mathrm{GeV}}
\def\TeV{\mathrm{TeV}}
\def\tL{{\widetilde L}}

\newcommand{\eqn}[1]{eq.\,(\ref{#1})}
\newcommand{\neqn}[1]{eqs.\,(\ref{#1})}
\newcommand{\fig}[1]{figure\,\ref{#1}}
\newcommand{\figs}[1]{figures\,\ref{#1}}
\newcommand{\tab}[1]{table\,\ref{#1}}
\newcommand{\sct}[1]{section\,\ref{#1}}
\newcommand{\scts}[1]{sections\,\ref{#1}}
\newcommand{\app}[1]{appendix\,\ref{#1}}

\def\refeq#1{\mbox{eq.\,\eqref{#1}}}
\def\refeqs#1{\mbox{eqs.\,\eqref{#1}}}
\def\reffi#1{\mbox{figure\,\ref{#1}}}
\def\reffitwo#1#2{\mbox{figures\,\ref{#1} and \ref{#2}}}
\def\reffis#1#2{\mbox{figures\,\ref{#1}--\ref{#2}}}
\def\refta#1{\mbox{table\,\ref{#1}}}
\def\reftatwo#1#2{\mbox{tables\,\ref{#1} and \ref{#2}}}
\def\reftas#1{\mbox{tables\,\ref{#1}}}
\def\refse#1{\mbox{section\,\ref{#1}}}
\def\refsetwo#1#2{\mbox{sections\,\ref{#1} and \ref{#2}}}
\def\refses#1{\mbox{sections\,\ref{#1}}}
\def\refapp#1{\mbox{app.\,\ref{#1}}}
\def\citere#1{\mbox{ref.\,\cite{#1}}}
\def\citeres#1{\mbox{refs.\,\cite{#1}}}

\newcommand{\rcut}{\ensuremath{r_{\mathrm{cut}}}}
\newcommand{\ww}{\ensuremath{W^+W^-}}
\newcommand{\wz}{\ensuremath{W^\pm Z}}
\newcommand{\wpz}{\ensuremath{W^+Z}}
\newcommand{\wmz}{\ensuremath{W^-Z}}
\newcommand{\z}{\ensuremath{Z}}
\newcommand{\w}{\ensuremath{W}}

\newcommand{\abbrev}{}
\newcommand{\llog}{\text{\abbrev LL}}
\newcommand{\nll}{\text{\abbrev NLL}}
\newcommand{\nnll}{\text{\abbrev NNLL}}
\newcommand{\lo}{\text{\abbrev LO}}
\newcommand{\nlo}{\text{\abbrev NLO}}
\newcommand{\nnlo}{\text{\abbrev NNLO}}
\newcommand{\nlonll}{\nlo\plus\nll}
\newcommand{\nnlonnll}{\nnlo\plus\nnll}
\newcommand{\qcd}{{\abbrev QCD}}
\newcommand{\D}{\mathrm{d}}

\newcommand{\cme}{centre-of-mass energy}
\newcommand{\cmes}{centre-of-mass energies}

\newcommand\Tstrut{\rule{0pt}{3.0ex}}         
\newcommand\Bstrut{\rule[-1.5ex]{0pt}{0pt}}   

\interfootnotelinepenalty=10000

\newcommand\mlbl[1]{{\mbox{\footnotesize #1}}} 

\newcommand{\elle}{\ensuremath{\ell}}
\newcommand{\genllln}{\ensuremath{\elle\elle\elle\nu}}
\newcommand{\llln}{\elle'^\pm{\nu}_{\elle^\prime} \elle^-\elle^+}
\newcommand{\mllln}{\ensuremath{m_{\llln}}}
\newcommand{\ptllln}{\ensuremath{p_{T,\llln}}}

\setlength{\tabcolsep}{5pt}

\makeatletter
\patchcmd{\@sect}{#8}{\boldmath #8}{}{}
\let\ori@chapter\@chapter
\def\@chapter[#1]#2{\ori@chapter[\boldmath#1]{\boldmath#2}}
\makeatother

\newcommand{\gev}[1]{$\unit{#1}{\giga\electronvolt}$}
\newcommand{\tev}[1]{$\unit{#1}{\tera\electronvolt}$}

\newcommand{\gevm}[1]{\unit{#1}{\giga\electronvolt}}
\newcommand{\tevm}[1]{\unit{#1}{\tera\electronvolt}}

\newcommand{\half}{$\frac{1}{2}$}

\newcommand{\ddk}[1]{\frac{d^d k_{#1}}{(4\pi)^d}}
\newcommand{\sidenote}[1]{\todo[noline]{#1}}

\newcommand\calo[1]{{\cal O}\hspace{-0.2em}\left(#1\right)}

\newcommand{\cala}{{\cal A}}
\newcommand{\bbH}{\ensuremath{b\bar{b}H}}
\newcommand{\ccH}{\ensuremath{c\bar{c}H}}
\newcommand{\bbtoH}{\ensuremath{b\bar{b}\rightarrow H}}
\newcommand{\qqtoH}{\ensuremath{q\bar{q}\rightarrow H}}
\newcommand{\ttH}{\ensuremath{t\bar{t}H}}
\newcommand{\bbphi}{\ensuremath{b\bar{b}\phi}}
\newcommand{\yt}{\ensuremath{y_t}}
\newcommand{\ytsq}{\ensuremath{y_t^2}}
\newcommand{\yb}{\ensuremath{y_b}}
\newcommand{\ybsq}{\ensuremath{y_b^2}}
\newcommand{\ybyt}{\ensuremath{y_b\, y_t}}

\newcommand{\phiF}{\Phi_{\text{F}}}
\newcommand{\phirad}{\Phi_{\text{rad}}}
\newcommand{\phiFJ}{\Phi_{\text{FJ}}}
\newcommand{\obs}{\mathcal{O}}
\newcommand{\barphiF}{\bar{\Phi}_{\text{F}}}
\newcommand{\barphiFJ}{\bar{\Phi}_{\text{FJ}}}
\newcommand{\dpwg}{\Delta_{\text{pwg}}}
\newcommand{\lpwg}{\Lambda_{\text{pwg}}}
\newcommand{\ptF}{{p_{\text{\relscale{0.77}T,$F$}}}}
\newcommand{\phiFJJ}{\Phi_{\text{FJJ}}}
\newcommand{\meF}{M_{\scriptscriptstyle\mathrm F}}
\newcommand{\meFJ}{M_{\scriptscriptstyle\mathrm FJ}}

\newcommand{\phiQQF}{\Phi_{\text{X}}}
\newcommand{\phiQQFJ}{\Phi_{\text{XJ}}}
\newcommand{\phiQQFJJ}{\Phi_{\text{XJJ}}}
\newcommand{\cflavF}{c_{\text{X}}}
\newcommand{\cflavFJ}{c_{\text{XJ}}}
\newcommand{\cflavFJJ}{c_{\text{XJJ}}}

\newcommand{\msb}{\overline{\text{MS}}}

\newcommand{\muIR}{{\mu_{\text{\relscale{0.77}S}}}}
\newcommand{\muRuno}{{\mu_{\text{\relscale{0.77}R,1}}}}
\newcommand{\muRdue}{{\mu_{\text{\relscale{0.77}R,2}}}}
\newcommand{\muIRuno}{{\mu_{\text{\relscale{0.77}S,1}}}}
\newcommand{\muIRdue}{{\mu_{\text{\relscale{0.77}S,2}}}}
\newcommand{\muRunotwo}{{\mu^2_{\text{\relscale{0.77}R,1}}}}
\newcommand{\muRduetwo}{{\mu^2_{\text{\relscale{0.77}R,2}}}}
\newcommand{\muIRunotwo}{{\mu^2_{\text{\relscale{0.77}S,1}}}}
\newcommand{\muIRduetwo}{{\mu^2_{\text{\relscale{0.77}S,2}}}}

%% file: sections/introduction.tex
\section{Introduction}

Since the discovery of a new scalar resonance with a mass of approximately 125~GeV~\cite{ATLAS:2012yve,CMS:2012qbp}, data collected at the Large Hadron Collider (LHC) have firmly established the existence of the Higgs boson as predicted by the Brout--Englert-Higgs mechanism responsible for electroweak symmetry breaking in the Standard Model (SM)~\cite{Higgs_theory1,Higgs_theory2}. Within this framework, the Higgs boson couplings to all elementary particles (including the Higgs boson itself) are universally determined by their corresponding masses. As a result, precise measurements of Higgs boson couplings offer a stringent test of the SM and a sensitive probe of possible extensions of the electroweak sector.\\

A global analysis of Higgs boson production and decay rates, combining multiple production modes and a diverse set of final states, shows that the couplings to vector bosons and third-generation fermions are currently measured with an accuracy of roughly 10--15\%, and are in good agreement with SM expectations~\cite{CMS:combination2026,ATLAS-CONF-2025-006}. Despite the success of SM predictions across a wide range of energy scales, several aspects of the Higgs sector remain largely unexplored. In particular, the Higgs boson self-coupling is still not constrained at a level close to the SM prediction~\cite{atlas2026combinationatlascmssearches}, and direct information on second-generation Yukawa interactions is limited.\\

A detailed study of Higgs boson couplings in both decay and production processes is therefore interesting both in its own right and relevant to $\PH\PH$ searches. Such an approach not only probes the fundamental structure of the SM, but also provides complementary handles for reducing theoretical and experimental uncertainties. The bottom-quark Yukawa coupling $y_b$ represents a particularly interesting case. Even though its magnitude is significantly smaller than those of the top quark and the electroweak gauge bosons, the decay $\PH \rightarrow \PQb\PAQb$ dominates the total width of a 125~GeV Higgs boson, largely due to phase-space factors. However, experimental determinations of $y_b$ are limited by overwhelming backgrounds with genuine or mistagged $\PQb$-jets and the large branching ratio $\mathrm{BR}(\PH \to \PQb\PAQb)$ makes a precision extraction of partial widths challenging. For these reasons, Higgs boson production modes that depend directly on the $\PQb\PAQb\PH$ coupling provide an attractive and complementary strategy to probe $y_b$\footnote{Dedicated searches for \bbH{} production have been performed~\cite{CMS:2024goa}, although no experimental observation has been reported so far.}.\\

Beyond third-generation Yukawas, a central open question is the strength of the charm-quark Yukawa coupling $y_c$. The smaller value of the charm coupling compared to $y_b$ suppresses both direct and indirect probes, thus resulting in extremely challenging measurements. Therefore, $y_c$ is currently only very weakly constrained~\cite{CMS:nature,ATLAS:2022vkf}, and remains compatible with values that substantially deviate from the SM prediction, leaving ample room for potential new-physics effects. From an experimental perspective, a robust determination of $y_c$ has far-reaching implications, as it would improve our knowledge of the Higgs boson interactions with second-generation quarks and provide critical input for global fits of Higgs boson couplings and effective field theory parameters. This context underscores the importance of identifying processes in which the dependence on $y_c$ is enhanced and theoretically well controlled.\\

Associated production of a Higgs boson with a heavy-flavour (HF) quark provides a direct and complementary approach to these problems. In the case of charm (bottom) quarks, the process $\Pp\Pp \rightarrow \PH+\PQc$ $(\PQb)$ (or, more generally, $\PH + \PQc(\PQb)$-jet production) is directly sensitive to the $\PQc\PAQc\PH$ ($\PQb\PAQb\PH$) coupling, alongside the dominant $y_t$-induced contribution. The leading order Feynman diagrams for the inclusive $\PH+\PQc$ $(\PQb)$ process, both $y_t$ and $y_c$ ($y_b$) dependent, are shown in Fig.~\ref{fig:FD_ggh}-\ref{fig:FD_b}. These channels offer a production-based probe of $y_c$ $(y_b)$, independent of the challenges that limit the precision of $\PH \rightarrow f\bar{f}$ decay studies. A particular advantage of this method, compared to the search for $\PH \rightarrow f\bar{f}$, lies in the fact that we probe $y_c$ and $y_b$ at production, \textit{via} the interaction with a quark from the abundant $\Pg+\PQc$ or $\Pg+\PQb$ initial state, allowing the reconstruction of the Higgs boson from its clean decay modes (such as $H \rightarrow \gamma\gamma$\ or $H\rightarrow \PW\PW$), thus reducing the problem of the non-Higgs background. Moreover, requiring a single HF-tagged jet in the final state allows to employ high-purity tagging algorithms in order to further reduce backgrounds. Finally these contributions are sensitive to the sign of the coupling through the interference with $y_t$ dependent contribution to the same final state. However, the downsides of these channels are the presence of large reducible and irreducible backgrounds due to the dominant Higgs boson production modes~\cite{Deutschmann:2018avk, Pagani:2020rsg, Manzoni:2023qaf}, which makes it extremely challenging to test the SM hypothesis for the heavy-quark Yukawa coupling.\\

In this paper, we address the challenges associated with the simulation of $\PH+$HF processes and provide practical recommendations for modelling single ($\PQc\PH$ and $\PQb\PH$) and double ($\PQc\PAQc\PH$ and $\PQb\PAQb\PH$) heavy-flavour associated Higgs boson production. In Section~\ref{sec:theory}, we review the theoretical frameworks used to describe $\PH+$HF production, discussing the different gauge-invariant contributions, the available factorisation approaches, and the current state-of-the-art predictions. While Higgs boson production via the bottom-quark Yukawa interaction has been extensively studied, the charm-Yukawa interaction remains comparatively unexplored~\cite{Brivio:2015fxa}. We therefore devote particular attention to the charm sector, first quantifying the impact of interference and internal quark-loop contributions in Section~\ref{sec:interference}. The setup of the QCD simulations is described in Section~\ref{sec:setup}, while the phenomenological results are presented in Section~\ref{sec:pheno}, focusing on key Higgs boson and charm-jet observables.

%% file: sections/simulation.tex
\section{Theoretical framework}\label{sec:theory}

Beyond the phenomenological motivations discussed above, the associated production of a Higgs boson with heavy-flavour quarks is theoretically interesting in its own right because of the possible factorisation approaches and the different gauge-invariant contributions.

\subsection{Flavour schemes}
Regarding the factorisation approaches, as is well known, the presence of a heavy quark in the hard process allows one to formulate the perturbative computation in different flavour schemes (FS), each appropriate to a specific kinematic regime. For Higgs boson production in association with charm (bottom) quarks, two factorisation approaches are typically employed.
A scheme where the charm (bottom) quark mass is not neglected, called massive, or three- (four-)flavour scheme ---3FS (4FS)---, and a scheme, called massless or 4FS (5FS), where the heavy quark is treated as a light quark neglecting its mass in the short-distance interaction. These schemes differ in their treatment of the heavy-quark mass and in the organisation of the perturbative expansion, and therefore provide complementary information.\\

Focusing on the case of the charm quark as a first example, in the 3FS, the physical charm-quark mass $m_c \sim 1.5$~GeV is retained in the short-distance cross section. The scheme is appropriate when the hard scale $Q$ characteristic of the process is of the same order of $m_c$, and that both
are scales at which QCD can be considered perturbative, i.e.
\begin{equation}
\max\!\left(\frac{\Lambda_{\rm QCD}}{Q},\, \frac{\Lambda_{\rm QCD}}{m_c}\right) \ll \frac{m_c}{Q},
\label{eq:4fs_condition}
\end{equation}
so that power-suppressed terms $(m_c/Q)^n$ are not negligible. In this setup, $\PQc$ quarks are treated as massive final-state objects and do not appear in the initial state. At leading order, the relevant partonic channels for inclusive $\PQc\PAQc \PH$ production are:
\begin{equation}
\Pg\Pg \to \PQc\PAQc \PH, \qquad \PQq\bar{\PQq} \to \PQc\PAQc \PH,
\end{equation}
with $\PQq=\PQu,\PQd,\PQs$. The 3FS provides full kinematic information on the final-state $\PQc$ quarks, making it naturally suited for event selections requiring the presence of one or more c-tagged jets.\\

At any fixed order $k$, however, the 3FS prediction contains logarithmic terms of the form $\alpha_s^k \log^k(m_c/Q)$ (arising from initial-state splittings), which are harmless when $Q \sim m_c$ but may spoil the perturbative convergence when $Q \gg m_c$. In the regime 
\begin{equation}
\max\!\left(\frac{\Lambda_{\rm QCD}}{Q},\, \frac{m_c}{Q}\right) \ll \frac{\Lambda_{\rm QCD}}{m_c},
\label{eq:5fs_condition}
\end{equation}
the logarithms $\log(m_c/Q)$ become large. 
The all-order resummation of these large logarithms can be achieved by employing a massless scheme, the 4FS in the case of the charm quark. In this scheme, the charm quark is treated as a massless parton on the same footing as the light quarks. Hence, it has a non-zero partonic distribution function inside the proton, and the leading-order process is the simple $2\to 1$ channel:
\begin{equation}
\PQc\PAQc \to \PH.
\end{equation}
More precisely, the massless scheme is known as Variable-Flavour Number Scheme (VFNS) since the heavy-quark effects are introduced at leading power via space-like threshold matching conditions in PDFs. The resummation of such contributions, known as collinear logarithms, is achieved through DGLAP evolution of the non-intrinsic heavy-quark PDF, as discussed for example in~\cite{Forte:2010ta}. Furthermore,
given the smaller final-state multiplicity, the computation of higher order corrections is easier in a massless scheme, and typically reaches higher orders that are not available in a massive scheme. However, the conceptual simplicity of the 4FS comes with limitations: information on final-state $\PQc$ quarks only arises at higher orders, and the absence of a mass regulator leads to unphysical behaviour in $\PQc$-tagged observables when the tagging transverse momentum is too low. These issues do not affect 3FS computations.\\

A conceptually analogous situation arises for Higgs boson production in association with bottom quarks. Here the relevant flavour schemes are the four-flavour scheme (4FS), where the bottom quark is treated as massive and does not appear in the initial state, and the five-flavour scheme (5FS), where bottom is treated as a massless parton. The 4FS is appropriate when the characteristic scale of the process is of the order of the bottom-quark mass $m_b \sim 5$~GeV; in this case, final-state bottom quarks can be treated perturbatively and the leading processes resemble those of the charm-quark 3FS case. Conversely, when $Q \gg m_b$ the 5FS provides a better description by resumming logarithms of $m_b/Q$ through the bottom PDF~\cite{Maltoni:2012pa,Harlander:2011aa}. As with the charm case, 5FS computations are simpler but lack explicit final-state bottom quarks at leading order, and fixed-order 5FS predictions may develop unphysical behaviour at low $p_T$ when the bottom is treated as massless. Most of the existing works focusing on the massive and massless schemes and the effect of the factorisation choice are performed for the bottom quarks. In particular, the fixed-order massless predictions have been known for a long time at NNLO in QCD~\cite{harlander_2003} and recently the next-to-NNLO (N3LO) corrections have been computed in Ref.~\cite{Duhr:2019kwi}. Monte Carlo generators that perform NNLO+PS simulations in the massless schemes are known using the \minnlo{} and \GENEVA{} methods~\cite{Monni:2019whf,Monni:2020nks,Alioli:2013hqa,Gavardi:2025zpf}. In the case of the massive scheme, the NLO QCD corrections have been computed in~\cite{dittmaier:2003ej,dawson:2003kb,Wiesemann:2014ioa,Jager:2015hka}, NLO in electro-weak theory and complete-NLO predictions have been performed in Ref.~\cite{Zhang:2017mdz,Pagani:2020rsg}, while recently the NNLO-accurate predictions~\cite{Biello:2024pgo} have been performed using the \minnlo{} method, extended for heavy-quark phenomenology in Refs.~\cite{Mazzitelli:2021mmm,Mazzitelli:2024ura}, with a two-loop approximation based on the high-energy limit of the amplitude~\cite{Mitov:2006xs,Badger:2024mir}. We refer to the report~\cite{biello2025} and references therein for details on modelling the \bbh{} process.\\

Several matching schemes have been developed to systematically combine 4FS and 5FS information into a single fully-inclusive cross-section, including the methods of Refs.~\cite{Harlander:2011aa,Forte:2010ta,Maltoni:2012pa,Bonvini:2015pxa,Forte:2015hba,Forte:2016sja,Forte:2019hjc,Duhr:2020kzd}. 
The same technology can, in principle, be extended to the charm-quark case; however, a detailed discussion of consistent flavour-scheme matching for the $y_c^2$ contributions lies beyond the scope of this work. Here, we instead present a first phenomenological combination of massive- and massless-flavour-scheme predictions in complementary $\PQc$-jet fiducial regions, postponing a fully consistent flavour-scheme matching to future studies.\\

Both massive and massless predictions are substantially improved by matching to parton showers. In particular, even the $\mathcal{O}(\alpha_s^0)$ 4FS (5FS) process $\PQc\PAQc \to \PH$ ($\PQb\PAQb \to \PH$) produces realistic charm-tagged (bottom-tagged) final states once backward evolution and hadronization are taken into account. For Higgs boson plus charm (bottom) production, the 3FS (4FS)  and 4FS (5FS) are expected to give compatible results in regions of phase space where neither mass effects nor resummation effects dominate.

\subsection{Gauge-invariant contributions}

\begin{figure}[h!]
    \centering
    \begin{tabular}{cc}
      \raisebox{2mm}{\includegraphics[width=0.33\textwidth, trim=0 0 4pt 10pt, clip]{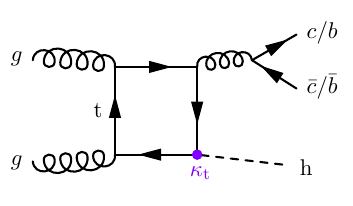}}
      \includegraphics[width=0.33\textwidth, trim=0 0 4pt 12pt, clip]{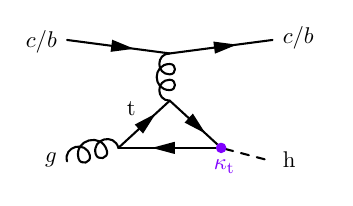}
    \end{tabular}
    \caption{Leading Feynman diagrams that contribute to $pp \rightarrow \PH+\text{HF}$ that do not depend on neither $\kappa_c$ nor $\kappa_b$ (where $\kappai=y_i/y_i^{\text{SM}}$). Purple dots correspond to vertices proportional to the top Yukawa coupling. Even if these processes are formally of higher order in $\alpha_S$, they represent the leading contribution to the $\PH+\text{HF}$ cross section, since they are enhanced by the presence of $\kappa_t$. Left: leading diagram contributing in a massive FS, right: leading diagram contributing in a massless FS.}
    \label{fig:FD_ggh}
  \end{figure}

\begin{figure}[h!]
      \centering
    \begin{tabular}{cc}
      \includegraphics[width=0.27\textwidth]{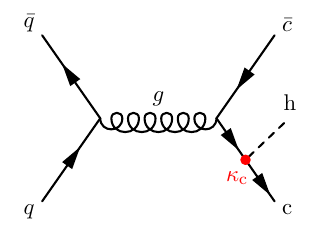}    
      \includegraphics[width=0.27\textwidth]{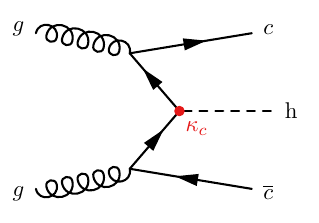} \\
      \includegraphics[width=0.27\textwidth]{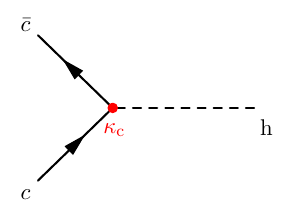}  
      \includegraphics[width=0.27\textwidth]{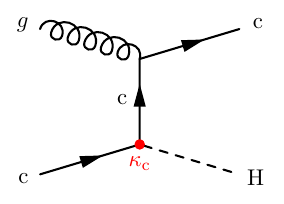} 
    \end{tabular}
      \caption{Tree-level \kappac dependent production of H+c in the 3FS (upper row) and 4FS (lower row). The diagrams in the lower row enter the 4FS calculation: the left diagram at LO, and the right diagram as the real-emission contribution at NLO. The \kappac component of \hpc  production is referred to as \ch and \cch in the text, depending on the FS.}
      \label{fig:FD_c}
  \end{figure}

\begin{figure}
      \centering
    \begin{tabular}{cc}
      \includegraphics[width=0.27\textwidth]{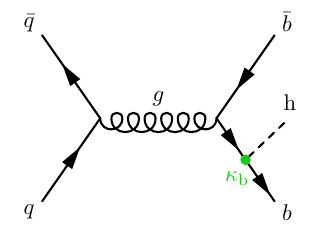}    
      \includegraphics[width=0.27\textwidth]{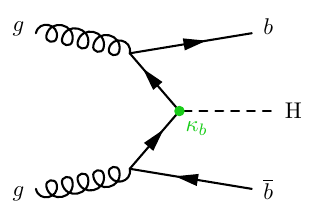}  \\
      \includegraphics[width=0.27\textwidth]{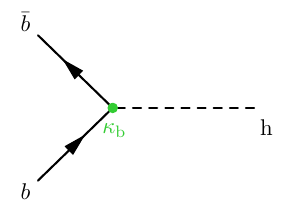}  
      \includegraphics[width=0.27\textwidth]{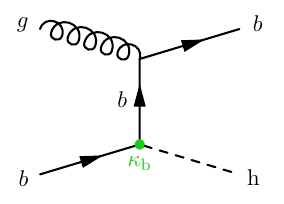}  
    \end{tabular}
      \caption{Same as Figure~\ref{fig:FD_c}, for the \kappab dependent production of H+b in the 4FS (upper row) and 5FS (lower row). The \kappab component of \hpb  production is referred to as \bh and \bbh in the text, depending on the FS.}
      \label{fig:FD_b}
  \end{figure}

The Feynman diagrams for the primary processes contributing to $\PH+\text{HF}$ production in the SM are shown in Figure~\ref{fig:FD_ggh},~\ref{fig:FD_c} and~\ref{fig:FD_b} in both FSs (massive and massless). The dominant production mechanism is gluon-gluon fusion, where final-state $\PQc$- and $\PQb$-quarks originate from processes such as gluon splitting or from a real HF quark extracted from the proton PDFs (Fig.~\ref{fig:FD_ggh}). These mechanisms do not provide a contribution proportional to the charm and bottom couplings, since the Higgs boson here is produced through a loop of top quarks. Additional contributions arise from charm- (\cch) and bottom-quark (\bbh) fusion as well as associated \ch and \bh production (Fig.~\ref{fig:FD_c},~\ref{fig:FD_b}).\\

The total cross section for the \hpc process can be expressed, in the case of a massive flavour scheme, as:
\begin{equation}
\mathrm{d}\sigma
= y_t^{2}\,\alpha_s^{4}\!\left(\Delta^{(0)}_{y_t^{2}} + \mathcal{O}(\alpha_s)\right)
+ y_t y_c\,\alpha_s^{3}\!\left(\Delta^{(0)}_{y_c y_t} + \mathcal{O}(\alpha_s)\right)
+ y_c^{2}\,\alpha_s^{2}\!\left(\Delta^{(0)}_{y_c^{2}} + \mathcal{O}(\alpha_s)\right)
\label{eq:xsec_hc}
\end{equation}
while for \hpb production it becomes:
\begin{equation}
\mathrm{d}\sigma
= y_t^{2}\,\alpha_s^{4}\!\left(\Delta^{(0)}_{y_t^{2}} + \mathcal{O}(\alpha_s)\right)
+ y_t y_b\,\alpha_s^{3}\!\left(\Delta^{(0)}_{y_b y_t} + \mathcal{O}(\alpha_s)\right)
+ y_b^{2}\,\alpha_s^{2}\!\left(\Delta^{(0)}_{y_b^{2}} + \mathcal{O}(\alpha_s)\right)
\, 
\label{eq:xsec_hb}
\end{equation}
and it does not scale trivially with the Yukawa couplings. $\Delta^{(0)}_{X}$ denotes the leading-order (LO) contribution to each term. The first contributions in Eq.~\ref{eq:xsec_hc} and~\ref{eq:xsec_hb} corresponds to the diagrams in Figures~\ref{fig:FD_c} and~\ref{fig:FD_b} respectively, where the Higgs boson directly couples to the charm and bottom quarks. The last terms correspond to the loop-induced gluon–fusion contribution, proportional to the top quark Yukawa coupling $y_t$ squared and are shown in Figure~\ref{fig:FD_ggh} (where the Yukawa couplings are expressed in terms of $\kappai=y_i/y_i^{\text{SM}}$). Despite being formally of higher order in $\alpha_s$, they are enhanced by the $y_t/y_{b/c}$ ratio, which appears in the loop, and results in being the dominant contribution to both cross sections: at 13~TeV they are approximately a factor of 10 and a factor of 2 larger than the $y_c^2$ and $y_b^2$ components, respectively. Finally, the middle terms correspond to the interference of the purely $y_{c/b}^2$ and $y_t^2$ terms and are usually a subleading contribution to the total cross sections.

We stress that some of these contributions cannot be captured within the massless scheme. In particular, as discussed in Sec.~\ref{sec:interference}, the interference contribution vanishes due to chirality conservation when the heavy quark is treated as massless in the short-distance dynamics. Moreover, even for the remaining gauge-invariant contributions, power corrections in the heavy-quark mass are neglected in the VFNS.

\subsection{Treatment of the Yukawa coupling}\label{sec:YukawaMSbar}
In contrast to the $\PQt\PAQt\PH$ case, where predictions with the Yukawa coupling renormalised in the on-shell (OS) scheme provide a viable option, the relatively small masses of the bottom and charm quarks strongly favour the use of the $\overline{\text{MS}}$ scheme for the corresponding Yukawa couplings~\cite{H1980}. Renormalizing the fermion--Higgs-boson coupling introduces logarithmic terms involving the fermion to Higgs boson mass ratio into the radiative corrections. In hadronic Higgs boson decays such as $\PH \rightarrow \PQq\overline{\PQq}(\Pg)$, these QCD corrections can surpass the LO decay rate, highlighting the limitations of a naive application of perturbation theory.\\

In the context of a mass-independent renormalization prescription, as in the case of QCD corrections, one can define a ``\textit{running mass}'' alongside the running coupling constant. QCD corrections are then more reliably obtained by replacing the quark pole mass in the lowest-order decay rate with its running mass~\cite{H1980}. 
Given that $m_c = 1.29$~GeV is relatively small and too close to $\Lambda_{\mathrm{QCD}}$ for perturbative calculations to be reliable, we suggest to set the starting value for the running mass $\bar{m}_c$ at $Q=3$~GeV, with $\bar m_c(\text{3 GeV})=0.986\,\text{GeV}$~\cite{Higgs_handbook}. The evolution of the coupling at the hard scale is performed following Eq.~(13) of Ref.~\cite{Harlander:2003ai} as in the bottom-quark scenario, where the evolution of the Yukawa coupling is correlated with the evolution of the strong coupling constant between the heavy-quark mass scale and the hard scale. The numerical values of the strong coupling are obtained through a running with consistent number of active flavours starting from the value at the $Z$-boson mass provided by the PDF set. For the central prediction of the charm Yukawa coupling, a four-loop evolution with four (three) active flavours is employed in the massless (massive) case. For scale variations, a two-loop evolution (for NLO+PS simulations) and a three-loop evolution (for \minnlo{} simulations) are used, consistently with the perturbative accuracy of the hard component. For \hpb{} simulations, instead, the bottom-quark mass is set to $\bar m_b(\text{4.18 GeV})=4.18\,\text{GeV}$, and the running is performed with five active flavours following the LHCHWG recommendation~\cite{Higgs_handbook,biello2025}.

\section{Leading-order study of interference and internal-loop effects}\label{sec:interference}

In this section, we present a leading-order study of non-standard contributions that depend on the charm-quark Yukawa coupling, namely the interference between the charm- and top-quark Yukawa interactions and the genuine charm-loop contribution to Higgs boson production via gluon fusion. The study is performed within the \MGvATNLO framework (\MADGRAPH in the following)~\cite{MadGraph,aMCatNLO,Frederix_2018}. The purpose of this analysis is to demonstrate that these effects are subleading with respect to the dominant charm-fusion ($y_c^2$) and top-induced ($y_t^2$) contributions, and can therefore be neglected as a first approximation. Dedicated higher-order studies of these more intricate contributions are left to future work. Such effects are expected to become increasingly relevant at higher luminosities, particularly in differential measurements.

\subsection{Interference effects}

The magnitude of the different terms for the production of \bbh was studied in detail in Ref.~\cite{Marius,biello2025}. In order to quantify the different contributions in Eq.~\ref{eq:xsec_hc} in the case of the charm quark, we performed LO parton-level simulations in the 3FS, where the charm quark is treated as massive, using \MADGRAPH. 
In the 4FS, with $m_c = 0$, the interference gives a vanishing contribution due to chirality conservation. The inclusive results are consistent with those reported in Ref.~\cite{Brivio:2015fxa} and are reported in Table~\ref{tab:3FSinterference} as a function of \kappac. The simulation distinguishes between four components: ``Full'', including all contributing diagrams, ``Yukawa'', including only the $y_c^2$-dependent terms, ``ggF'', including only the non-$y_c$-dependent diagrams and the interference term between the ``Yukawa'' and ``ggF'' components. Among the three subcomponents, the term proportional to $y_t^2$ contributes most significantly to the total \hpc cross section, while the $y_c^2$ dependent provides sensitivity to Higgs-charm coupling and contributes at the $\mathcal{O}(10\%)$ level. The interference term is smaller, contributing at the $\mathcal{O}(1\%)$ level for $y_c \sim y_c^{\text{sm}}$. \\

\begin{table}[ht!]
    \centering
    \begin{tabular}{llll}
    \toprule
    $\kappa_c$ & 1 & 3 & 10 \\
    \midrule
    Full        & 0.208 & 0.366 & 2.215 \\
    ggF         & 0.191 & 0.191 & 0.191 \\
    Yukawa      & 0.021 & 0.186 & 2.056 \\
    Interference & -0.003 & -0.009 & -0.030 \\
    Interf./Yukawa & -0.142 & -0.048 & -0.014 \\
    \bottomrule
    \end{tabular}
    \caption{Integrated LO cross sections [pb] for $pp \rightarrow \PH \PQc\bar{\PQc}$ in the 3FS scheme. The fiducial phase-space is defined requiring a leading charmed jet with a minimum transverse momentum $p_T(\PQc) > 25$~GeV.}
    \label{tab:3FSinterference}
\end{table}

NLO corrections to the $y_c \cdot y_t$ interference can be computed by building a suitable model for \MADGRAPH, as it has been done for the case of the bottom quark in Ref.~\cite{Deutschmann:2018avk}.
However, due to its small numerical impact and taking into account the theoretical uncertainties on the simulation (PDFs, scales, and flavour schemes) and the limited sensitivity of current analyses targeting \hpc~\cite{CMS_chgg2025,ATLAS_chgg2025} associated production, we choose, and suggest, to neglect the interference in the first approximation.~\footnote{
 A dedicated calculation of the interfrence at NLO exists~\cite{Bizon_2021}, which shows that NLO corrections
 do not alter the LO hierarchy among contributions.} This decision may need to be revisited in future studies, especially if the sensitivity of experimental constraints improves toward the $y_c \sim y_c^{\text{sm}}$ limit.\\

\begin{figure}[h!]
    \centering
    \includegraphics[width=0.49\textwidth]{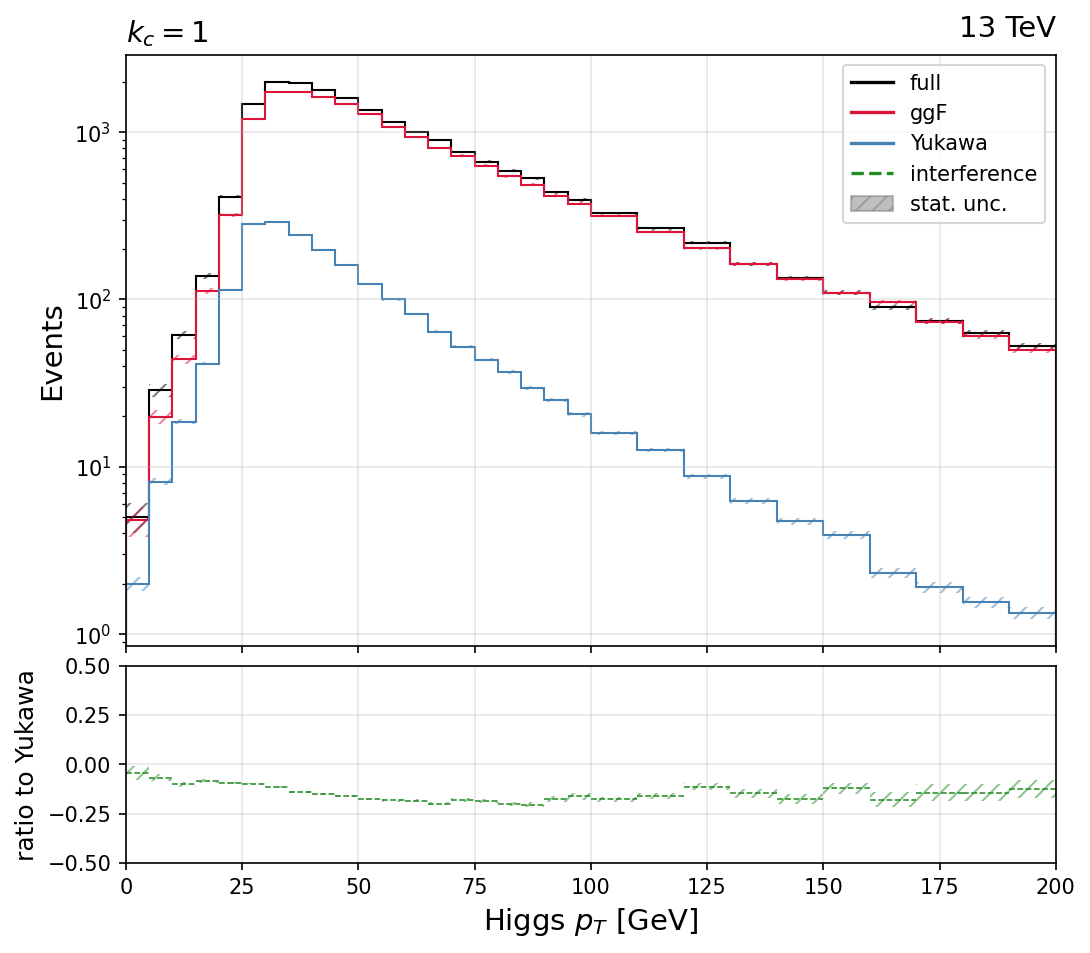}
    \includegraphics[width=0.49\textwidth]{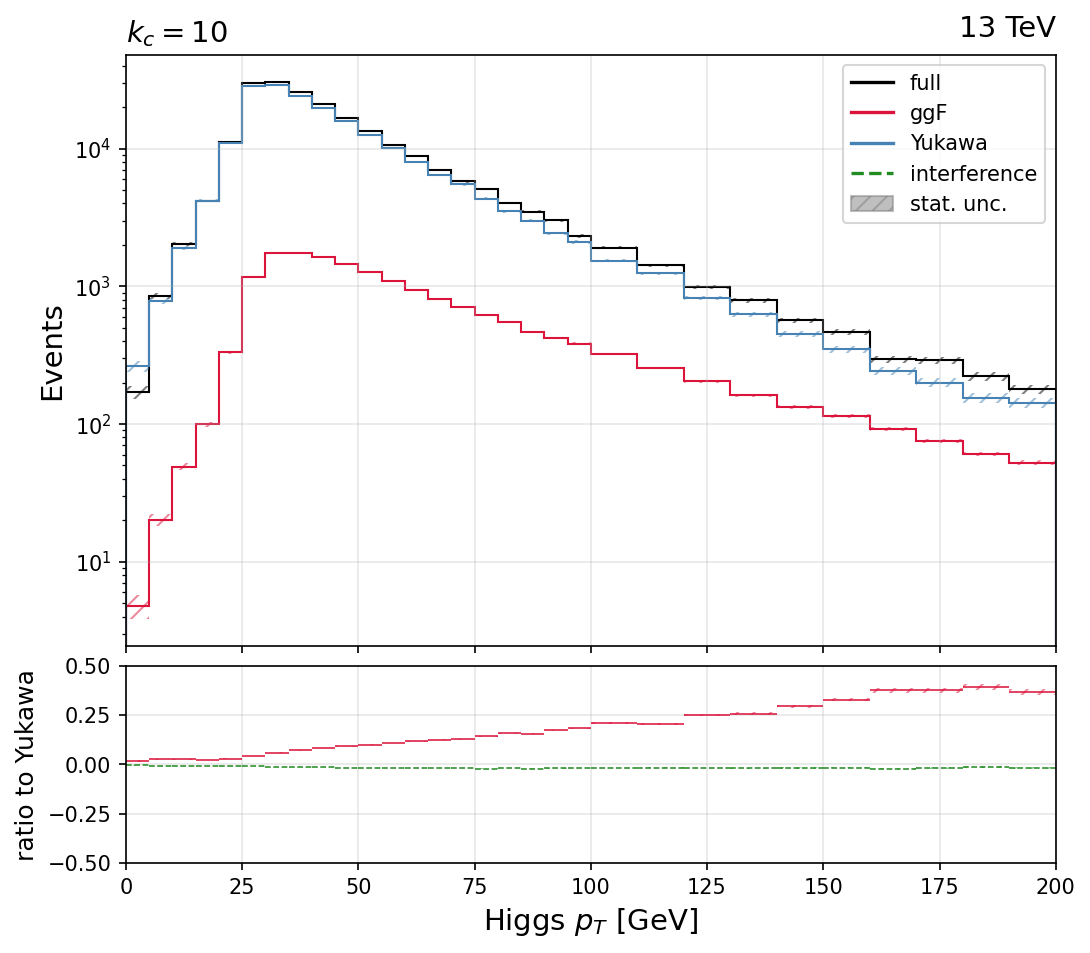}
    \caption{Differential cross section components for $\kappa_c = 1$ (left) and $\kappa_c = 10$ (right) as a function of Higgs boson $p_T$ in 3FS. "Full" includes all diagrams; "Yukawa" includes only $y_c^2$ terms; "ggF" includes non-$y_c$-dependent terms. The interference is defined as Yukawa $\times$ ggF.}
    \label{fig:4FSIntDiff}
\end{figure}

A further validation was performed to assess the differential stability of the interference term's magnitude. The objective was to evaluate the relative contribution of the interference with respect to the Yukawa dependent component across differential observables such as the transverse momentum of the Higgs boson.
Inclusive results for the massive scheme are presented in Table~\ref{tab:3FSinterference} as a function of \kappac, while differential results as a function of the Higgs boson \pt and \kappac are shown in Figure~\ref{fig:4FSIntDiff}.
The results show that the relative interference contribution remains approximately constant and small across the Higgs boson $\pt$ spectrum. Moreover, the interference relative contribution to the total cross section scales inversely with $\kappa_c$ ($\sim 1/\kappa_c$, for large $\kappa_c$). Therefore, for analyses targeting $\kappa_c > 1$, the $y_c\cdot y_t$ term becomes negligible compared to the dominant $y_c^2$ component, further supporting the decision of neglecting it at current sensitivity levels.

\subsection{Heavy-flavour loop contributions in gluon-fusion Higgs boson production}

\begin{figure}[ht!]
        \includegraphics[width=0.43\textwidth]{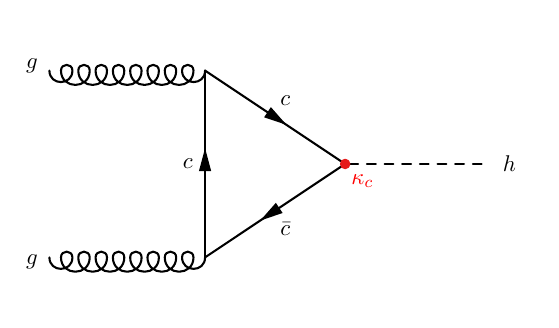}
        \centering
        \includegraphics[width=0.35\textwidth]{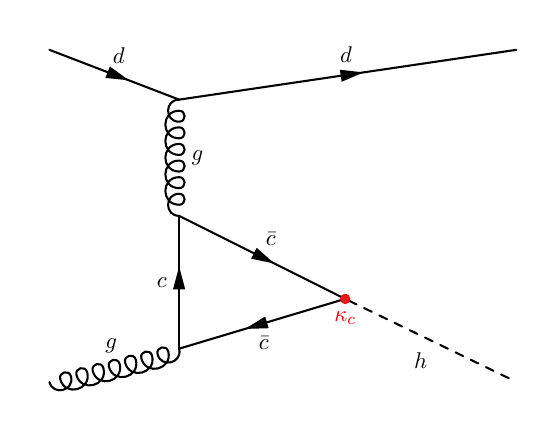}
        \centering
        \caption{Examples of loop induced diagrams with \ch coupling.}
        \label{fig:loops}   
\end{figure}

In addition to the aforementioned processes, there exist loop-induced contributions that couple the Higgs boson to charm quarks via closed-loop diagrams. These diagrams, illustrated in Figure~\ref{fig:loops}, are not automatically included in the simulations used in this analysis or in the \POWHEG{} and \MADGRAPH H+jets Monte Carlo samples, due to the fact that they do not have a Born level counterpart. 

We therefore perform a study on how to simulate these loop-induced terms, in order to assess the magnitude and relevance of these contributions in the context of an \hpc analysis. Some key points must be considered in this context:

\begin{enumerate}
\renewcommand{\labelenumi}{(\roman{enumi})}
    \item The loop-induced processes and the $\ch$ production simulated in the 4FS sample are entirely distinct. Since they do not share the same initial or final states, there is no interference, and their contributions could, in principle, be added in a straightforward manner.
    \item These simulations must also be performed using the $\MSbar$ renormalization scheme (see Sec.~\ref{sec:setup}).
    \item They dominantly contribute via the interference with $y_t$-induced loops, and distort the Higgs boson transverse-momentum spectrum~\cite{Bishara:2016jga}.
\end{enumerate}

Here, we focus on the contribution that
does not stem from interferences with the top-induced diagrams. The syntax used to generate the relevant processes in \MADGRAPH can be found in Appendix~\ref{appendix}.

All the resulting processes in this setup involve charm quarks neither in the initial nor final states. The charm Yukawa coupling enters through closed charm-quark loops. However, in the limit where the charm is treated as massless (as in 4FS), these loops vanish due to helicity conservation. Therefore, the simulation must be performed in the 3FS where the charm quark is massive.

Under this assumption, the total cross section is $\mathcal{O}(5~\fb)$. This value is further suppressed, by roughly an order of magnitude, when requiring the presence of a tagged charm object in the final state. Such tagging typically requires an additional gluon emission followed by a $g \to \PQc\bar{\PQc}$ splitting. From this study, we conclude that loop-induced contributions are negligible in the context of the signal modelling for the experimental analyses targeting these processes with the current precision.

\section{Methods and setup of QCD simulations for $\PH+\PQc$ production}\label{sec:setup}

Here, we focus on modelling the $y_c^2$ dependent component of $\PH+\PQc$ production in both the massless and massive schemes, neglecting the interference.
In the present section, we discuss the methods used to perform the parton-shower-matched simulations. In particular,
we employ \MADGRAPH to simulate both schemes at NLO QCD accuracy. For a better description of higher-multiplicity observables in the massless scheme, we also perform a simulation employing FxFx merging procedure~\cite{Frederix2012} including up to one extra jet at NLO. The NNLO+PS simulations have been performed using the \minnlo{} method in the \POWHEG{} framework. 

Parton showering, hadronization, and the underlying event are modelled by \PYTHIA{}8 (8.240 for NLO samples and 8.244 for \minnlo{} samples) with the same shower settings presented in Appendix~\ref{appendix}. In all simulations \PYTHIA is used to decay the Higgs boson to a pair of photons.

For all simulations we employed FS dependent PDF sets:
\begin{itemize}
    \item \textbf{3FS}: \texttt{NNPDF31\_nnlo\_pch\_as\_0118\_nf\_3,~lhaid=321900}.
    \item \textbf{4FS}: \texttt{NNPDF31\_nnlo\_as\_0118\_nf\_4\_mc\_hessian,~lhaid=325500}.
    \item \textbf{5FS}: \texttt{NNPDF31\_nnlo\_as\_0118\_mc\_hessian\_pdfas,~lhaid=325300}.
\end{itemize}

The values for the masses of various particles and couplings used in the different simulations are summarized here:

\begin{itemize}
    \item \textbf{Particle masses:} $m_c = 1.51\ (0)$~GeV in the 3FS (4FS),  $m_b = 4.92 \ (0)$~GeV in 4FS (5FS), $m_t=173$~GeV, $m_\tau=1.777$~GeV, $m_\PZ=91.18$~GeV, $m_\PW=80.0419$~GeV, $m_H=125$~GeV.
    \item \textbf{SM couplings:} $\alpha_{\mathrm{EW}}=132.507$, $G_F =1.166390\times10^{-5}~\textrm{GeV}^{-2} $, $\alpha_s(m_\PZ)=0.118$.
    \item \textbf{Yukawa couplings:} $y_c(3~\text{GeV})=0.986$ in \hpc simulations, while $y_c=0$ in the other cases; $y_b(m_b)=4.18$ in \hpb simulations, while $y_b=0$ in the other cases. The other Yukawa couplings are set to zero, in particular $y_t = y_e = y_\mu = y_\tau = 0$.
\end{itemize}

For the scale settings in the nominal results in these simulations we adopted $\mu_R=\mu_F=H_T/4$ and $\mu_\text{PS} = 0.5$ for all the NLO QCD simulations, while we used the internal scale choices of the \minnlo{} method with the Yukawa scale set to $\mu_R=m_H$ for NNLO+PS predictions. For a detailed description on how these values were selected we refer to Sec.~\ref{sec:pheno}.

\subsection{Simulations at NLO QCD with \MADGRAPH}

Neglecting the interference term and heavy-flavour internal loops, the simulation of the $\ch$ signal process (i.e., the $y_c^2$ term only) using \MADGRAPH is performed with NLO QCD precision. While this discussion uses the $\ch$ process as an example, all considerations apply \textit{mutatis mutandis} to $\bh$ production, with the corresponding shift in the flavour-scheme number, $3\to4$ and $4\to5$.\\

In this work, we present simulations in both the 3FS and 4FS. Our nominal predictions are obtained in the 4FS, as the 3FS is known to provide unreliable inclusive cross sections due to the presence of large logarithmic contributions that are not resummed. These effects are expected to be even more pronounced in the charm-quark case than in bottom-quark induced Higgs boson production, where they can be numerically significant depending on the specific scale choices. The massive 3FS simulations are therefore primarily employed to assess flavour-scheme uncertainties and to compare inclusive and differential observables.\\

\MADGRAPH can handle 3FS \cch{} and 4FS $\PQc\bar{\PQc} \rightarrow \PH$ production in a relatively straightforward manner. However, its default treatment of Yukawa couplings follows an on-shell renormalization scheme, which is suboptimal for Higgs boson production processes in association with bottom and lighter quarks. The $\overline{\textrm{MS}}$ renormalisation scheme can be implemented in the model, following procedures from previous studies on \bbh production~\cite{Marius}, and trough modifications that include a patched routine to perform the scale-dependent running of $m_c(\mu_R)$ following the prescription in Section~\ref{sec:YukawaMSbar}. Aside from these theoretical adjustments, the generation and computation of the $\PQc\bar{\PQc} \rightarrow \PH$ cross sections in both 3FS and 4FS proceed as in Ref.~\cite{MadGraph}, and documented in Appendix~\ref{appendix}. We stress that the UFO models used in the simulations include the appropriate UV counterterms for renormalizing the charm Yukawa coupling within the $\overline{\text{MS}}$ scheme.

\subsection{Merging ME and PS simulations with FxFx}

At next-to-leading order in the massless scheme, we employ the FxFx merging scheme to achieve a better precision, adding the description of an additional real emissions by keeping the NLO accuracy of the 4FS sample. This approach leverages on the Sudakov suppression in the PS below a merging scale $\mu_M$, while using ME predictions for hard emissions above $\mu_M$. In this work, FxFx merging is applied within the 4FS (and 5FS for \bbh) using \MADGRAPH, as the approach is not currently compatible with 3FS. We generate both inclusive Higgs boson production and Higgs boson+jet samples above a $\pt$ cut to regulate divergences. Each sample is matched to a parton shower and combined through a reweighting procedure to account for overlapping topologies. The generation commands for the FxFx-merged 4FS sample can be found in Appendix~\ref{appendix}.

\subsection{\minnlo{} simulations in the 4FS}\label{sec:minnlo}

Starting from the \minnlo{} implementation of the $\PQb\PAQb \to \PH$ process~\cite{Biello:2024vdh}, we perform NNLO event simulations of charm-quark fusion through the Yukawa interaction in 4FS. Unlike the predictions reported in~\cite{biello2025}, the present implementation is consistently formulated with four active flavours. In addition, the Yukawa coupling is set to the Standard Model charm-quark value according to the prescription described in Section~\ref{sec:YukawaMSbar}. The resulting generator therefore provides a fully consistent NNLO-accurate 4FS simulation, employing four-flavour parton distribution functions, a four-flavour running of the strong coupling constant, and short-distance matrix elements computed with only four active quark flavours.
In particular, contributions involving bottom quarks are consistently excluded. This includes both the double-real corrections with bottom quarks in the final state and the corresponding double-virtual contributions containing bottom-quark loops, which are removed through the explicit implementation of the $n_f$ dependence of the two-loop amplitude. \\

The simulations presented in this work follow the \minnlo{} prescription for the scale choices entering both the singular component of the $p_T$ spectrum and the regular contribution, which is evaluated at the same dynamical scale.\footnote{For bottom-quark fusion, evaluating the regular component at the hard scale was found to have a small impact in Ref.~\cite{Biello:2024vdh}.} The Yukawa coupling is renormalised at the Higgs-mass scale, and its scale variation is performed in a correlated way with that of the strong coupling. The resummation scale is chosen as $Q=K_Q m_H$ with $K_Q=0.25$, below which soft and soft-collinear logarithms are resummed through the \minnlo{} Sudakov form factor. We have verified that using $K_Q=0.5$ leads to very similar phenomenological predictions. The perturbative uncertainty is estimated from the standard 7-point variation of the renormalisation and factorisation scales, implemented through \POWHEG{} event reweighting.

The parton-shower simulation is performed with \PYTHIA~8, using the same settings as the \MADGRAPH{} predictions. In particular, multi-parton interactions are enabled, and the \texttt{lhaid = 325500} PDF set is employed for the backward evolution with four active flavours. We observe that a synchronised treatment of the backward evolution is important for the modelling of low-energy Higgs boson production in fiducial regions with tagged $c$-jets, leading to effects of up to $10$--$30\%$ in observables involving low transverse momenta of the Higgs boson relative to the leading charm jet.

\subsection{NLO+PS stitching of massless and massive predictions}\label{sec:stitched}
To overcome the differences between the heavy-flavour production cross sections computed in the two flavour schemes, a combination strategy would be beneficial. Achieving a fully consistent matching is theoretically non-trivial, since the organisation of the collinear mass logarithms differs between the two approaches~\cite{Gauld:2021zmq}, posing substantial challenges to the development of a consistent all-order matching framework~\cite{Gaggero_2022,Aglietti_2007,Cacciari_2002,Corcella_2004}.

In this section, we instead pursue a more experimentally oriented approach that can be easily implemented in data comparisons. It aims to leverage the main advantage of the massive scheme, namely that the kinematics of the heavy-flavour quarks is treated correctly with the full mass dependence retained, while relying on the simpler massless computation where possible.  We define a \textit{stitched} sample through a phase-space splitting based on the number of charm jets with $p_T>10$~GeV present in the event.\footnote{The $10$-GeV threshold is chosen lower than the usual analysis-level requirement of $25$~GeV in order to resolve the soft-charm region, where the two schemes differ most.} We take events with $N_{c\text{-jet}}=0$ from the 4FS-FxFx sample and events with $N_{c\text{-jet}}\geq1$ from the 3FS sample. The combined sample is normalised in two steps. First, each jet-multiplicity region (0 and $\geq$ 1 charm jets) is normalised to the fiducial cross section predicted by its originating sample for that region, corrected with a flat $K$-factor extracted from the NNLO/NLO ratio of the inclusive predictions. For the massive scheme, where no NNLO-accurate simulation of \cch{} is available, this factor is taken from \bbh{}, under the assumption that the higher-order correction between the bottom and charm case are similar. This assumption is further tested in Sec.~\ref{sec:NLOvsNNLO}, where the two are found to agree within~3\%. The full prediction is then rescaled to match the NNLO-accurate \cch cross section, in this work the integrated \minnlo{} result.

\section{Phenomenological results}\label{sec:pheno}

\subsection{NLO+PS study of scale choices and flavour-scheme differences at 13.0 TeV}

Several studies on \bbh production, including Refs.~\cite{Marius,biello2025}, have highlighted the sensitivity of NLO cross-section predictions to the choice of input parameters in \MADGRAPH, including renormalization ($\mu_R$), factorization ($\mu_F$), and parton-shower settings. We have thus studied the dependence of the cross section prediction for our \hpc (3FS and 4FS) and \hpb (4FS and 5FS) simulation as a function of these parameters to ensure a consistent and accurate prediction with our setup, comparing cross-section values and differential distributions under various scale configurations. In literature, for \bbh production two commonly used scale choices are:
\begin{equation}
    \mu_R = \mu_F = \frac{H_T}{4}, \qquad H_T = \sum_i \sqrt{m_i^2 + p_T(i)^2},
\end{equation}
and
\begin{equation}
    \mu_R = \mu_F = \frac{m_H + 2 m_b}{4}.
\end{equation}

In our analysis we opted to keep $\mu_R$ and $\mu_F$ equal, and to favour dynamic over fixed scales to avoid artifacts in the presence of high-$\pt$ partons in the ME calculation. To test these choices, we generated multiple samples varying the starting values of $\mu_R$ and $\mu_F$ and evaluated the impact of the parton-shower scale. The latter, loosely associated with the maximum emission hardness in the shower, was tuned via the parameter \texttt{shower\_scale\_factor} (PS scale factor).

Since the FxFx merging procedure can't be used together with the massive approach, we compared 3FS vs 4FS without FxFx merging for the \hpc samples and 4FS vs 5FS without FxFx merging for the \hpb samples. Table~\ref{tab:Samples} summarizes the tested configurations.\\

\begin{table}[ht!]
    \centering
    \begin{tabular}{ccc}
        \toprule
        Scale type & $\mu = \mu_R = \mu_F$ & PS scale factor \\
        \midrule
        dynamic & $H_T/8$ & $H_T/8$ \\
        dynamic & $H_T/4$ & $H_T/4$  \\
        dynamic & $H_T/4$ & $H_T/2$  \\
        dynamic & $H_T/2$ & $H_T/2$  \\
        dynamic & $3  \cdot H_T/4$ & $3  \cdot H_T/4$  \\
        dynamic & $H_T$ & $H_T$  \\
        fixed   & $(m_H + 2 m_c)/4$ & $H_T/4$ \\
        fixed   & $(m_H + 2 m_c)/4$ & $H_T/2$ \\
        \bottomrule
    \end{tabular}
    \vspace{0.3cm}
    \caption{3FS, 4FS and 5FS sample configurations tested.}
    \label{tab:Samples}
\end{table}

For the inclusive cross section, we can observe strong discrepancies between the two predictions. 
The observed difference becomes much smaller ($O(10-20\%)$) and compatible with the scale uncertainties, when the requirement of an analysis like fiducial phase space is applied, imposing the presence in the events of at least two photons from the Higgs boson decay with $p_T>25$~GeV and absolute pseudo-rapidity $|\eta| < 2.5$, and of a jet containing a D or B hadron with the same kinematic cuts. Furthermore, the massive FS presents a larger cross-section dependence on the selected value of the scales, with variations of the central prediction up to 70\%.\\

\begin{figure}[ht!]
    \centering
    \begin{minipage}{0.99\textwidth}
        \centering
        \includegraphics[width=\linewidth, trim=0 0 0 10pt, clip]{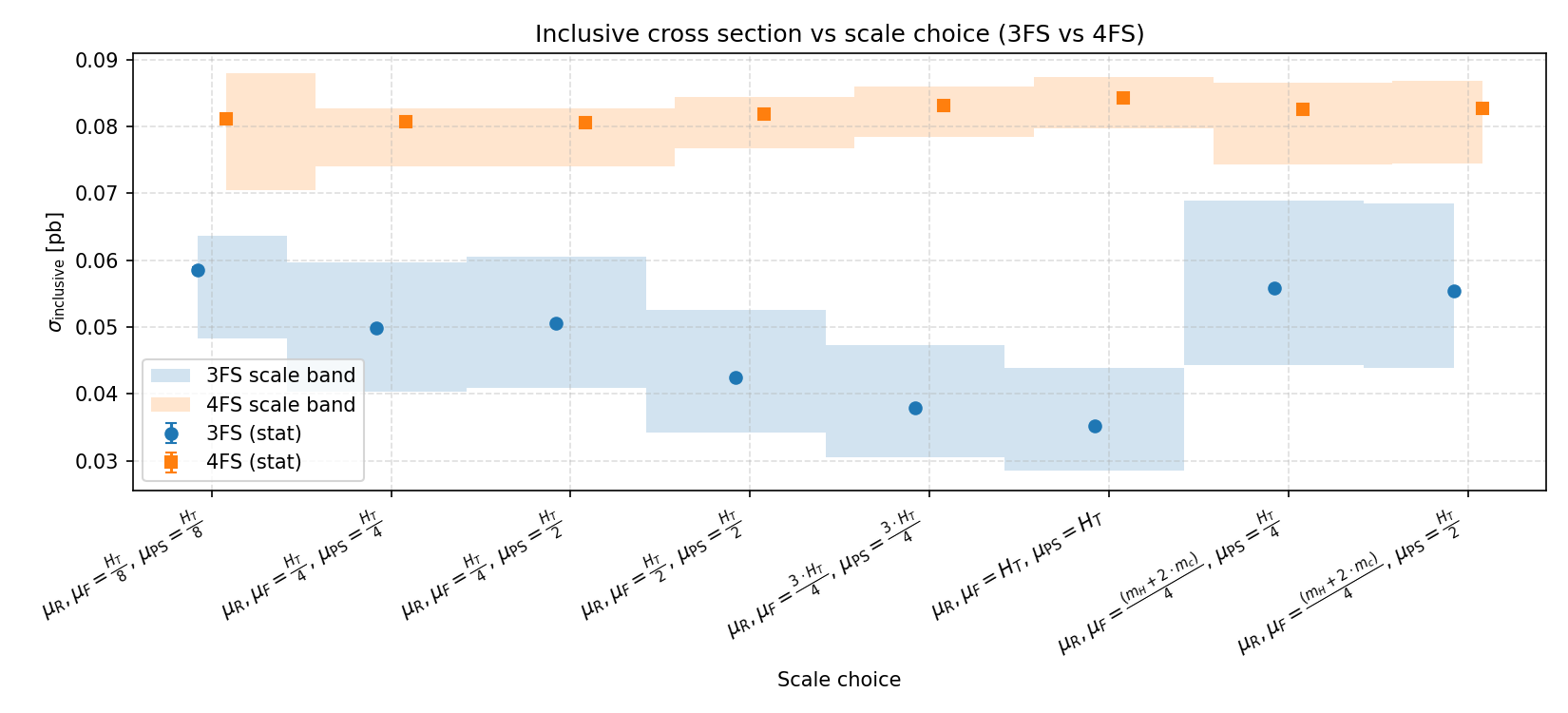}
    \end{minipage}\\
    \begin{minipage}{0.99\textwidth}
        \centering
        \includegraphics[width=\linewidth, trim=0 0 0 10pt, clip]{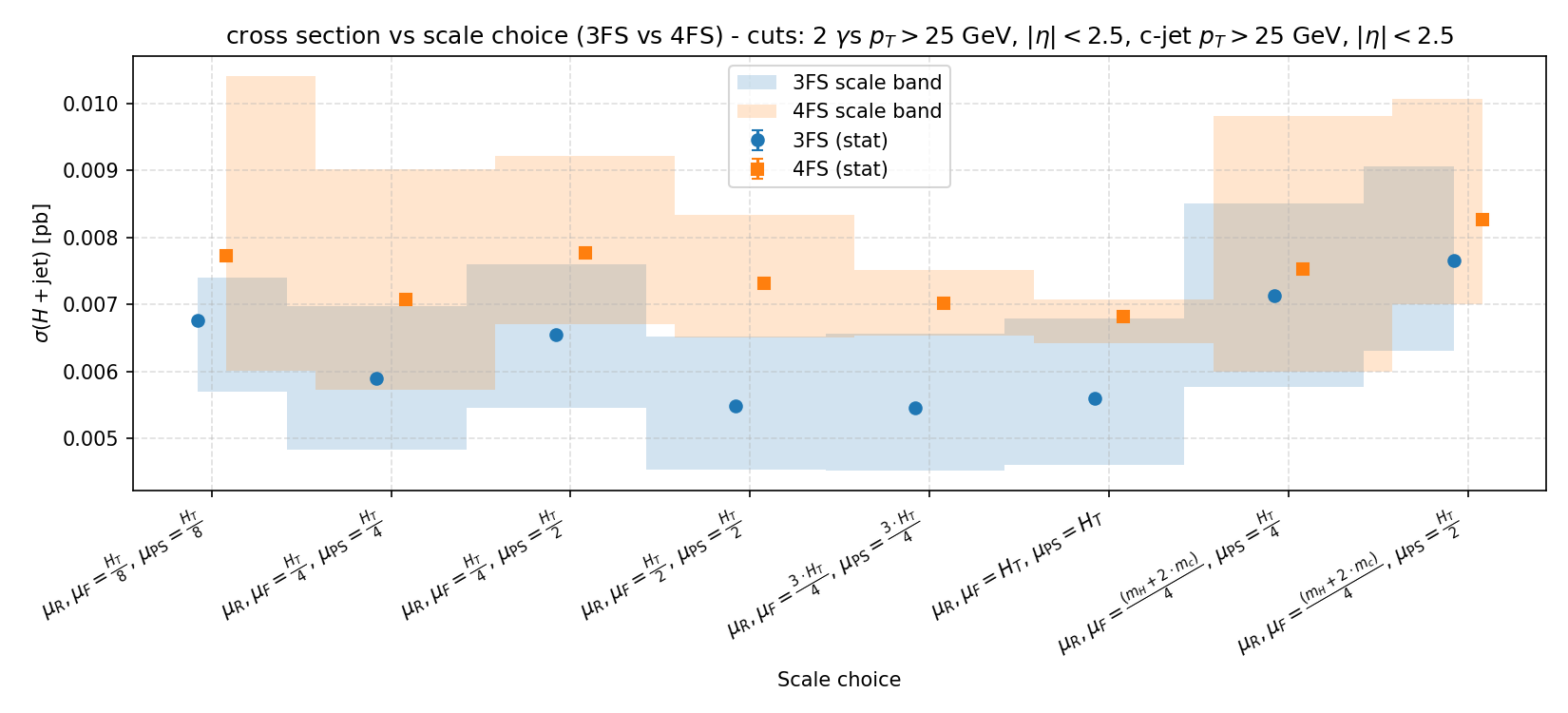}
    \end{minipage}
    \caption{The top plot shows the comparison of 3FS and 4FS inclusive \cch cross sections at 13.0 TeV w.r.t. the starting value of $\mu_{R/F/PS}$. The bottom plot describes the cross sections with fiducial cuts at 13.0 TeV. Shaded areas represent the 7 point scale variation envelope obtained by varying $\mu_R$ and $\mu_F$ values by a factor 2 up or down.}
          \label{fig:ch_xsec}
\end{figure}

\begin{figure}[ht!]
    \centering
    \begin{minipage}{0.99\textwidth}
        \centering
        \includegraphics[width=\linewidth, trim=0 0 0 10pt, clip]{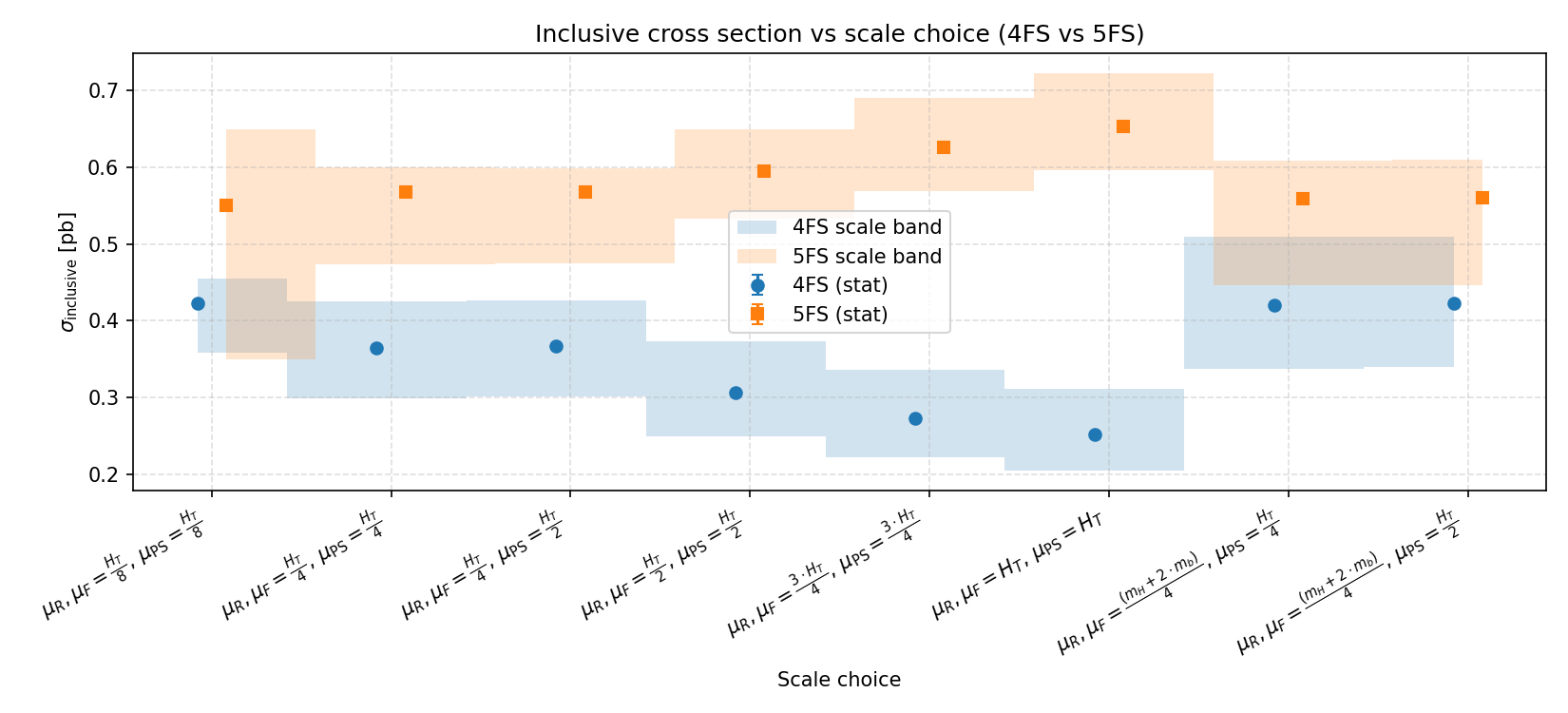}
    \end{minipage} \\
    \begin{minipage}{0.99\textwidth}
        \centering
        \includegraphics[width=\linewidth, trim=0 0 0 10pt, clip]{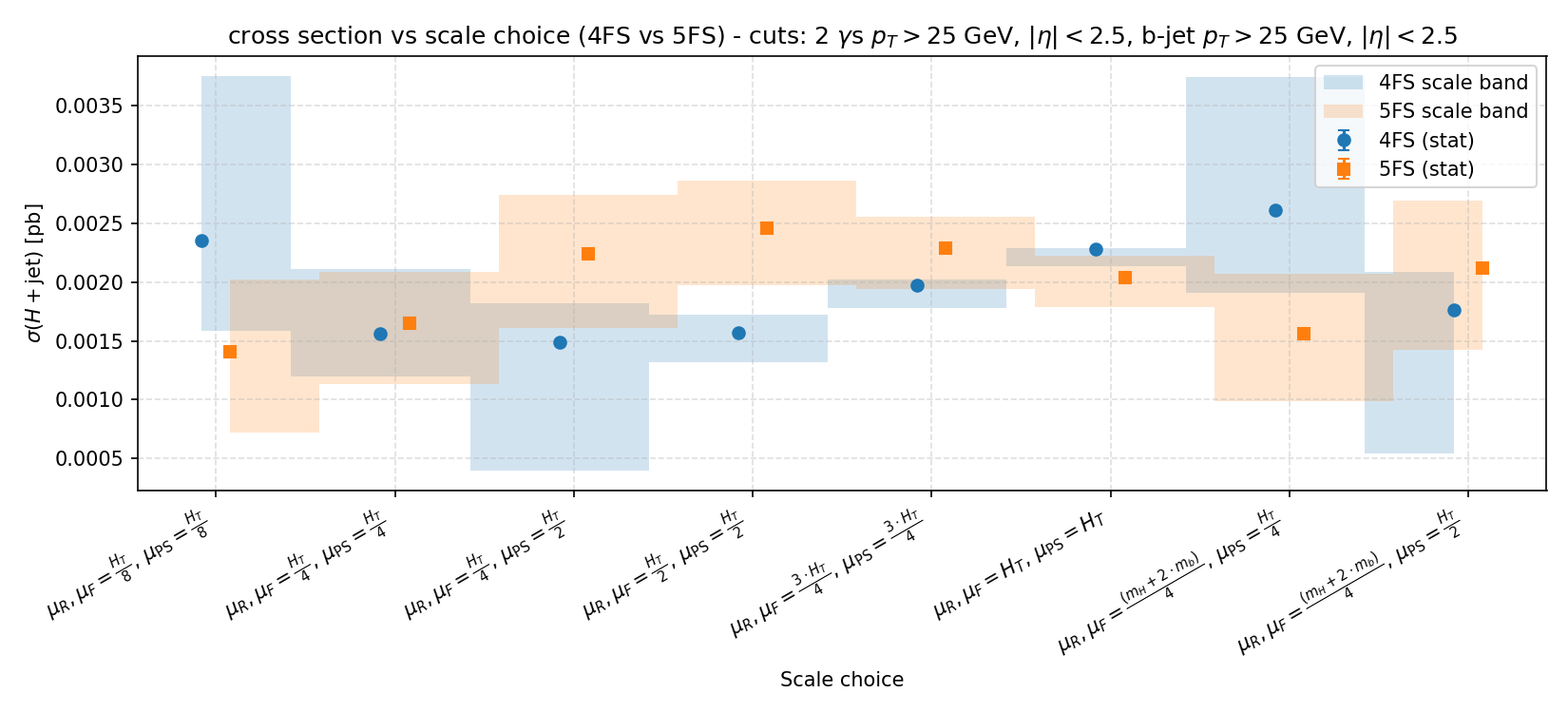}
    \end{minipage}

    \caption{Same as Fig.~\ref{fig:ch_xsec} for \bbh production. Top: 4FS vs 5FS inclusive \bbh cross section w.r.t. the starting value of $\mu_{R/F}$. Bottom: cross section  with fiducial cuts.}
    \label{fig:bh_xsec}   
\end{figure}

As figure of merit to decide the best configuration we adopted the difference between the cross-section predictions in the massive and massless FSs. This results in the choice of the initial values of $\mu_{R/F}\leq H_T/4$, corresponding to the first three bins in Fig.~\ref{fig:ch_xsec} and~\ref{fig:bh_xsec}, or the fixed scale choices, corresponding to the last two bins in Fig.~\ref{fig:ch_xsec} and~\ref{fig:bh_xsec}.

\subsection{FxFx merging at 13.0 TeV}
\begin{figure}[h!]
    \centering
    \includegraphics[width=0.49\textwidth]{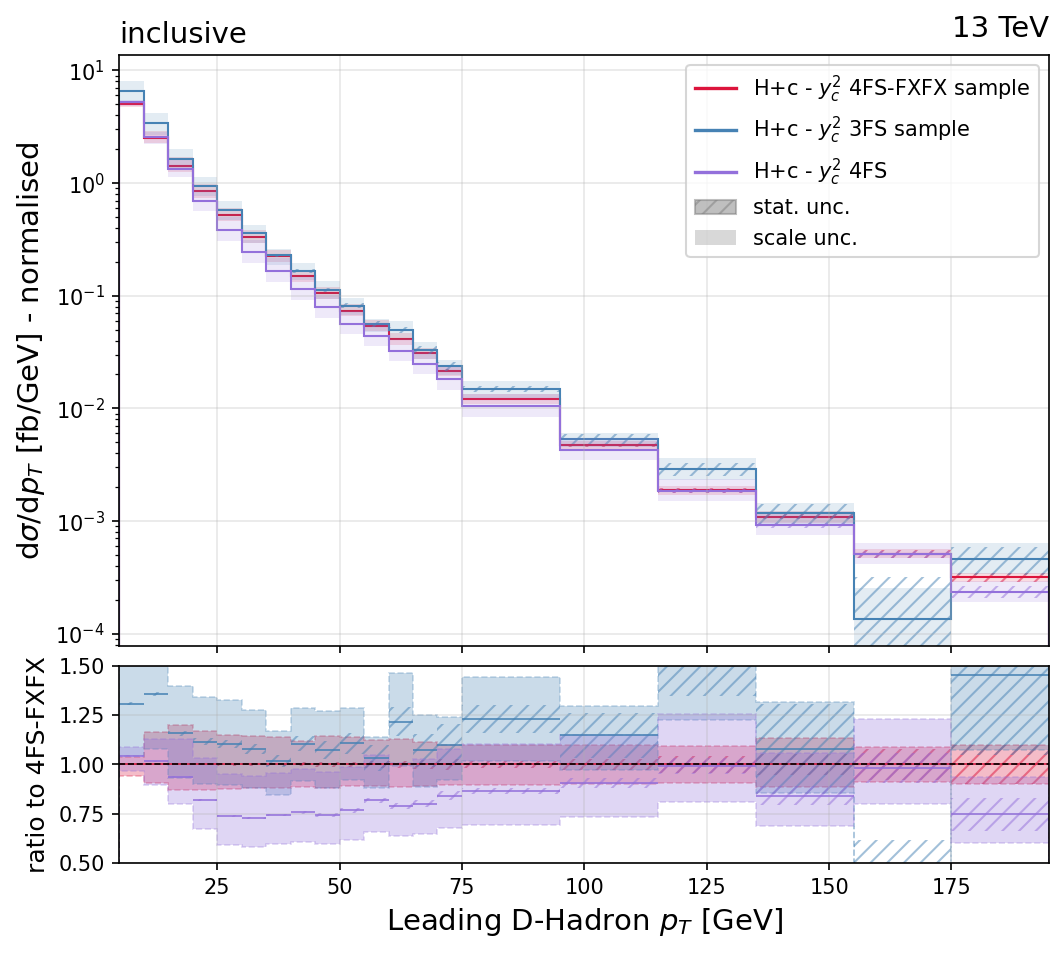}
    \includegraphics[width=0.49\textwidth]{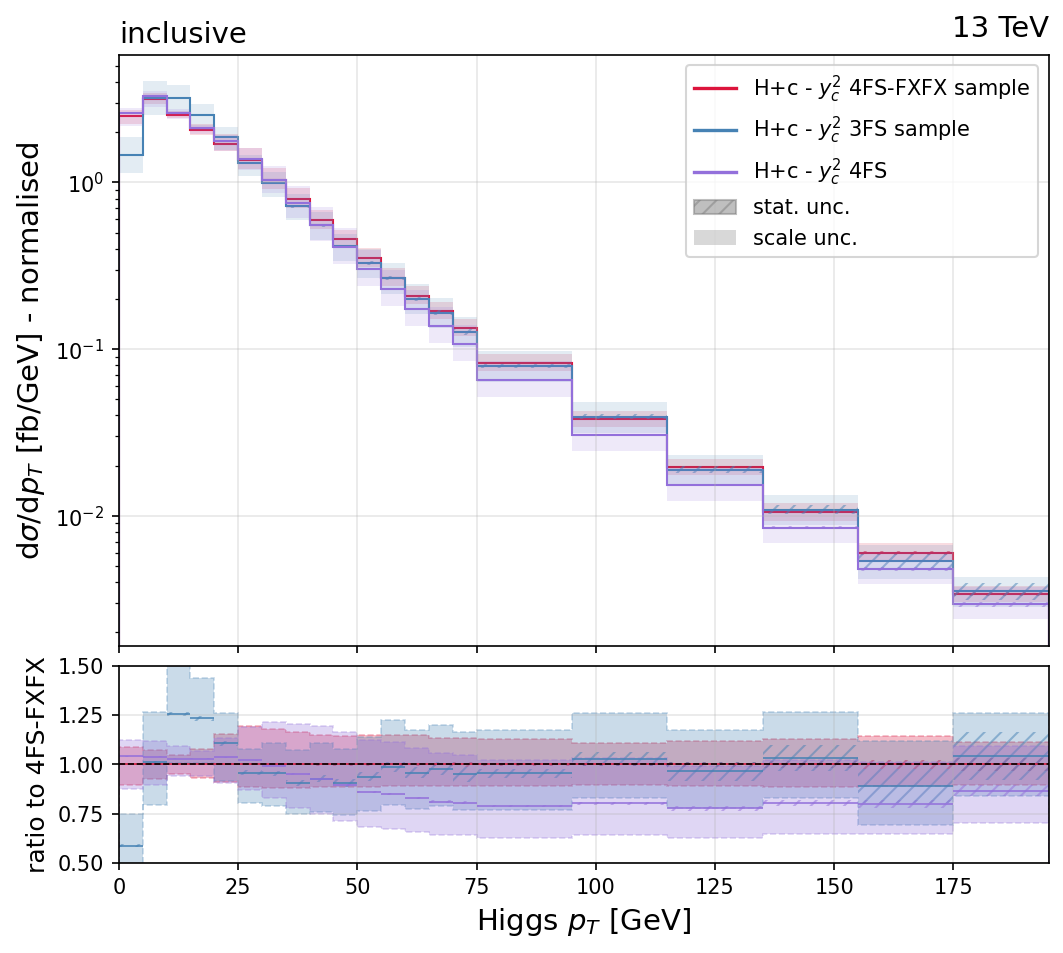}
    \includegraphics[width=0.49\textwidth]{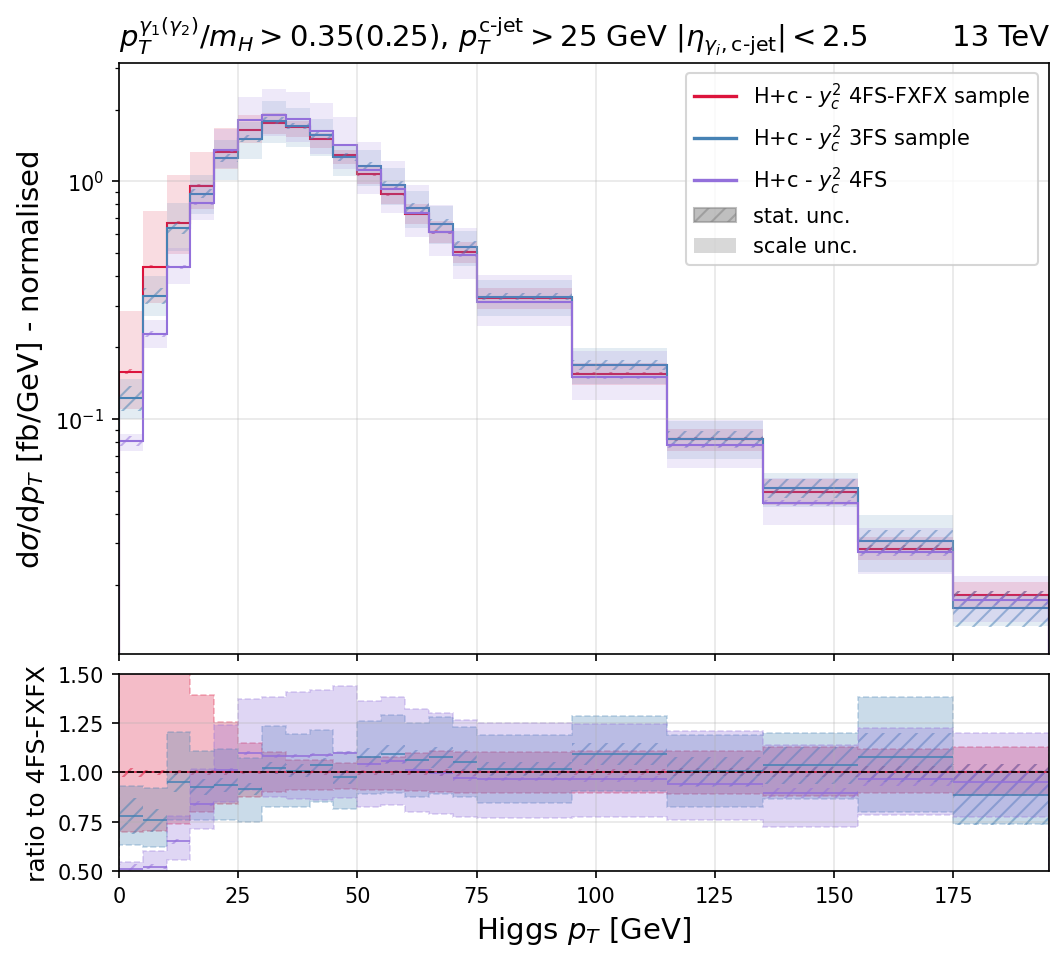}
    \caption{Hardest D-hadron and \PH $p_T$ spectrum obtained for different simulation setups: 3FS (blue), 4FS (purple), 4FS FxFx (red), shaded areas represent the 7-point scale variation envelope, while the error bars correspond to the statistical error of the simulation. The lower panel shows the ratio with respect to the 4FS simulation (with FxFx merging applied). The different samples are normalised to the same inclusive cross section.}
    \label{fig:DHad_pt_comparison}
\end{figure}

We now discuss the FxFx phenomenological results at 13.0 TeV and we compare them with the massless and massive predictions. To illustrate the predictions of the 4FS FxFx, 3FS, and 4FS, we consider the $p_T$ spectrum of the leading $D$-hadron and of the \PH boson, shown in Figure~\ref{fig:DHad_pt_comparison}. As expected, in the high momentum region, where mass effects are negligible, the 3FS and 4FS FxFx predictions yield comparable spectra in terms of shape with and without fiducial cuts (while still exhibiting a consistent difference in inclusive normalisation). In contrast, at low transverse momentum (below $p_{\mathrm{T}} \simeq 20~\mathrm{GeV}$), where mass effects play a role, a discrepancy is also observed at the level of shape of the \pt spectra for both the \PH boson and the leading D-hadron in the event. Furthermore, the 4FS prediction fails to reproduce the shape obtained in the 3FS and 4FS FxFx calculations at large momenta (even though the predictions are still compatible within uncertainties). This behaviour originates from the absence, at matrix-element level, of the hard second charm emission in the 4FS description. 

\subsection{Comparison of NLO+PS and NNLO+PS predictions at 13.6 TeV}\label{sec:NLOvsNNLO}

At the level of our simulation setup, we adopt the selected scale configuration for each flavour scheme ($\mu_R=\mu_F=H_T/4$, $\mu_{PS}=0.5$) and build a ``stitched'' sample through a phase-space splitting as defined in Section~\ref{sec:stitched}. We then compare the 3FS, 4FS with FxFx merging and the stiched samples with the \minnlo{} simulations in the massless scheme at 13.6 TeV. All simulations perform the Higgs boson decay into two photons during the parton-shower stage, based on the narrow-width approximation. We perform the comparison both inclusively and in fiducial phase spaces defined by the photon decay products and the leading $c$-jet, which is defined as the highest-$p_T$ jet containing at least one charm quark among those clustered with the anti-$k_T$ algorithm.\\

We start with a comparison of integrated cross-section numbers reported in Table~\ref{tab:minnloxsection}. At NLO level, the 4FS and 3FS predictions differ by nearly a factor of two (0.2216 vs 0.1248 fb in the inclusive case). This discrepancy is reduced in the fiducial region but remains sizeable, reflecting a large flavour-scheme uncertainty. The NNLO corrections in the 4FS, included through \minnlo{}, are negative and reduce the scale sensitivity with respect to the NLO result. In the $\gamma+c$ column the \minnlo{} prediction retains only NLO accuracy, which accounts for the correspondingly larger scale uncertainty. Reweighting the NLO predictions with the ccH$-\kappa_\mathrm{fact}$ factor, defined as the ratio between the NNLO and NLO \bbh{} cross sections in the corresponding scheme based on the results in Ref.~\cite{biello2025}, brings the 3FS and 4FS results into much closer agreement across all phase-space volumes. Also, it yields a 4FS prediction in good agreement with the dedicated \minnlo{} simulation for charm-quark fusion in the same scheme, thus validating this strategy to extrapolate the 3FS \cch prediction, not available at NNLO.\\

\begin{table}[ht!]
    \centering
    \renewcommand{\arraystretch}{1.5}
    \begin{tabular}{llll}
    \toprule
     & No cuts & $\gamma$ cuts & $\gamma+c$ cuts \\
    \midrule
    \minnlo{} (4FS)        & $0.1664^{+4.4\%}_{-2.7\%}$ & $0.1055^{+4.4\%}_{-2.8\%}$ & $0.0165^{+9.4\%}_{-8.7\%}$\\
    NLO (4FS)          & $0.2216^{+10.2\%}_{-7.0\%}$ & $0.1372^{+10.1\%}_{-6.9\%}$ & $0.0226^{+14.6\%}_{-11.9\%}$\\
    NLO (3FS)      & $0.1248^{+19.9\%}_{-19.3\%}$ & $0.0806^{+19.7\%}_{-19.2\%}$ & $0.0140^{+17.6\%}_{-17.7\%}$\\
    NLO+bbHfact (4FS) & $0.1690^{+10.2\%}_{-7.04\%}$ & $0.1046^{+10.1\%}_{-6.91\%}$ & $0.0172^{+14.6\%}_{-11.9\%}$ \\
    NLO+bbHfact (3FS) & $0.1630^{+19.9\%}_{-19.3\%}$ & $0.1054^{+19.7\%}_{-19.2\%}$ & $0.0182^{+17.6\%}_{-17.7\%}$\\
    \bottomrule
    \end{tabular}
    \caption{Integrated cross sections [fb] for $pp \rightarrow \PH(\gamma\gamma)\PQc\bar{\PQc}$ production at 13.6 TeV, including the Higgs boson branching factor. The first column gives the fully inclusive cross sections, while the second column reports the fiducial cross sections obtained by requiring $p_T^{\gamma_1}/m_H > 0.35$, $p_T^{\gamma_2}/m_H > 0.25$ and $|\eta^{\gamma_i}|<2.5$. The last column additionally requires at least one tagged $c$-jet with $p_T^{c\text{-jet}}>25$ GeV and $|\eta^{c\text{-jet}}|<2.5$. The NLO+PS predictions are shown both before and after applying the reweighting procedure based on the NNLO $\bbH$ results of Ref.~\cite{biello2025}.}
    \label{tab:minnloxsection}
\end{table}

\begin{figure}[h!]
    \centering
    \includegraphics[width=0.49\textwidth]{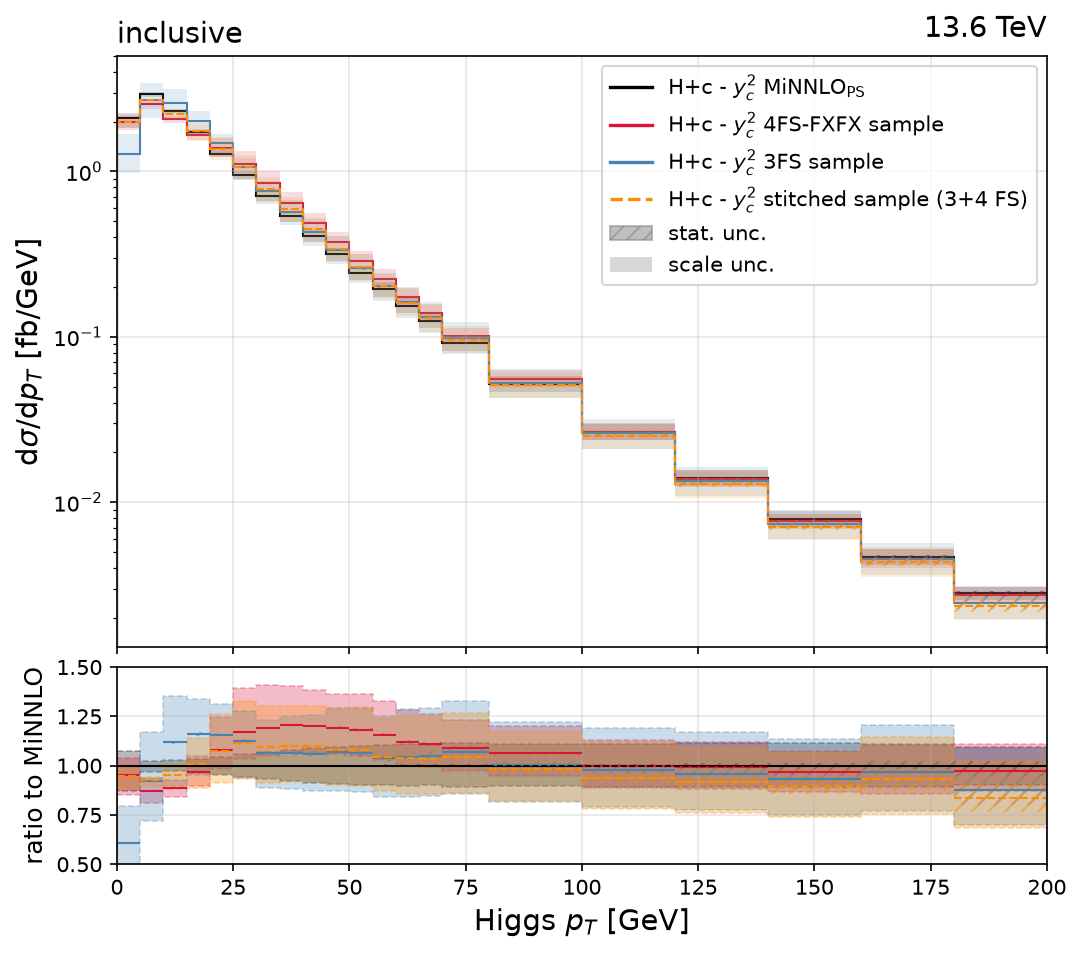}
    \includegraphics[width=0.49\textwidth]{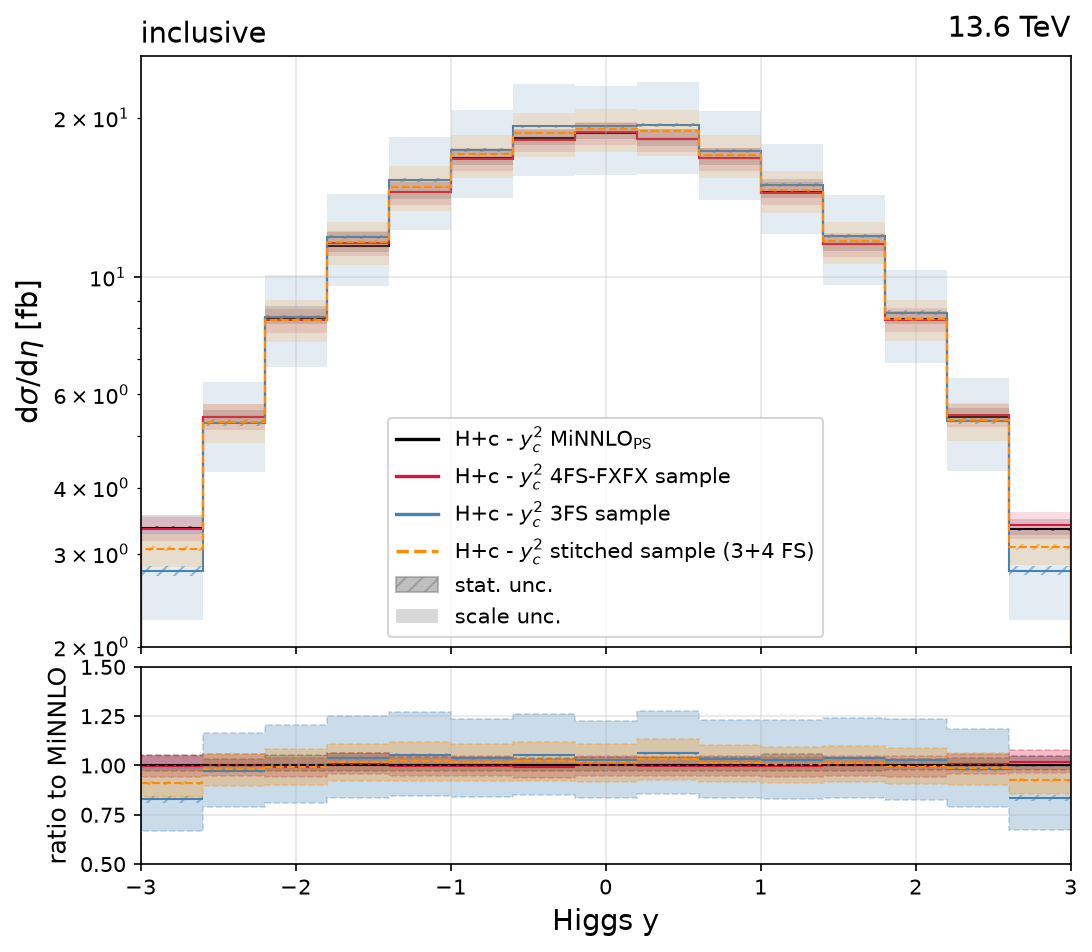}
    \includegraphics[width=0.49\textwidth]{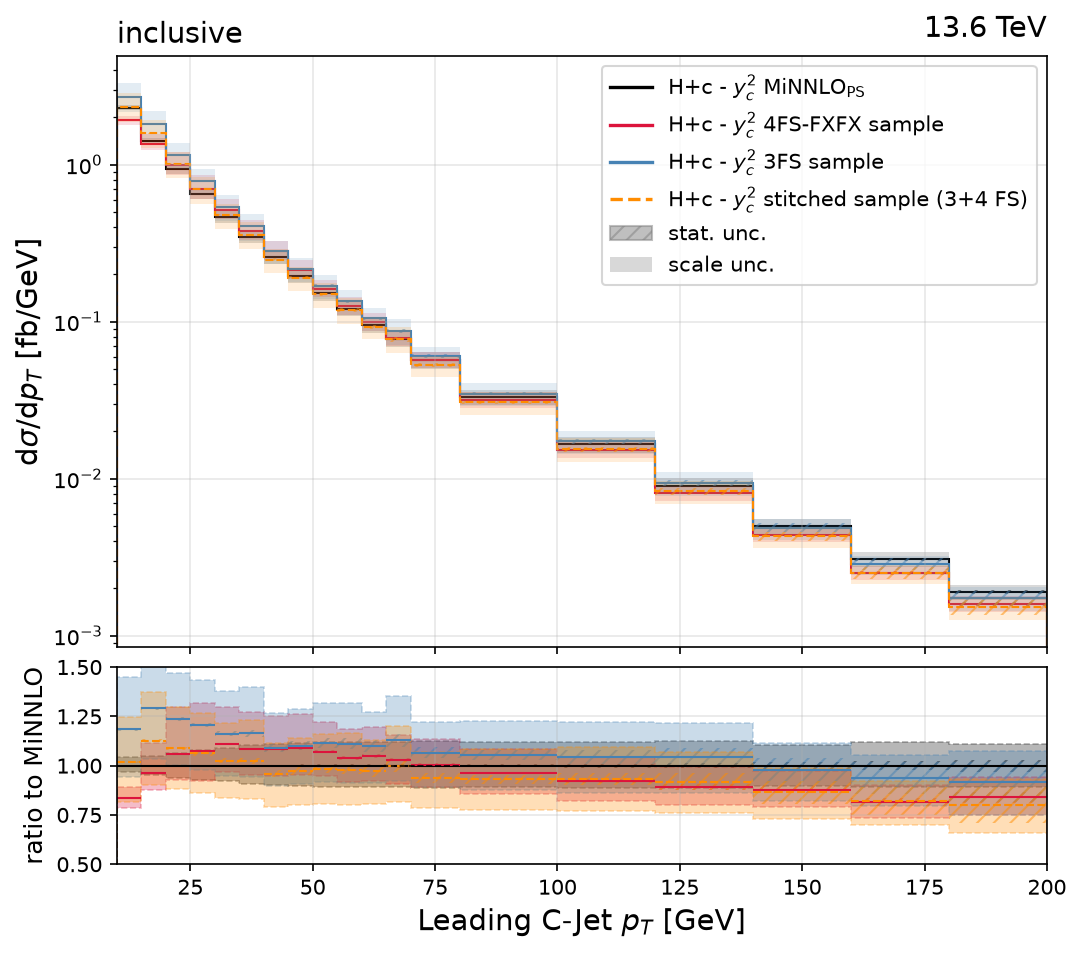}

    \caption{Differential $p_T(H)$, $y(H)$ and $p_T(\text{c-jet})$ cross section distribution comparison between the MiNNLOps sample and the 3FS, 4FS-FXFX and stitched samples. The three plot show the distributions without any fiducial cut on the \PH and a requirement on the $p_T$ of the c-jet to be greater than 10~GeV.}
    \label{fig:MiNNLO_inclusive}
\end{figure}

We continue with an inclusive analysis at differential level. We stress that the NLO+PS simulations have been reweighted to match the inclusive normalisation of the \minnlo{} predictions. Therefore, the purpose of this analysis is to assess the impact of NNLO corrections on differential distributions by comparing the shapes predicted at NLO+PS and NNLO+PS accuracy. In Figure~\ref{fig:MiNNLO_inclusive} we observe excellent agreement between the FxFx prediction and \minnlo{} in the high-$p_T^H$ tail, including the scale uncertainty, as expected since both predictions achieve the same perturbative accuracy. The stitched simulation improves upon the inclusive NLO+PS prediction, yielding a better description of the intermediate-$p_T^H$ region while increasing the associated uncertainty band. In the first low-$p_T^H$ bins, the stitched result also shows improved agreement with \minnlo{}. The residual shape differences are not unexpected, as \minnlo{} includes higher-order logarithmic contributions beyond the leading-logarithmic accuracy of the parton shower, improving the description of the low transverse-momentum region of the Higgs boson. At high transverse momentum, however, the stiched sample tends to follow more closely the 3FS prediction, which explains its deviation from both the \minnlo{} and FxFx results in that region.
In the second panel of Figure~\ref{fig:MiNNLO_inclusive}, we show the rapidity distribution of the Higgs boson. The FxFx prediction exhibits essentially the same shape as the \minnlo{} result across the entire spectrum. The massive-scheme prediction, and consequently the stitched sample, displays only mild shape effects at large rapidities, while in the central region it closely follows the behaviour of the massless predictions. The third panel of Figure~\ref{fig:MiNNLO_inclusive} shows the transverse-momentum distribution of the leading charm jet. This observable is NLO accurate in the massless simulations based on both the FxFx and \minnlo{} methods. The same accuracy is achieved in the massive scheme, where the full final-state charm-quark kinematics is available. We observe a similar shape for all four predictions, with the stitched result showing better agreement with the \minnlo{} prediction in the low-transverse-momentum region. However, we observe a small suppression of the FxFx prediction with respect to \minnlo{} in the high-$p_T$ tail of the spectrum, likely originating from differences in the matching procedures employed by the two approaches.\\

\begin{figure}[h!]
    \centering
    \includegraphics[width=0.49\textwidth]{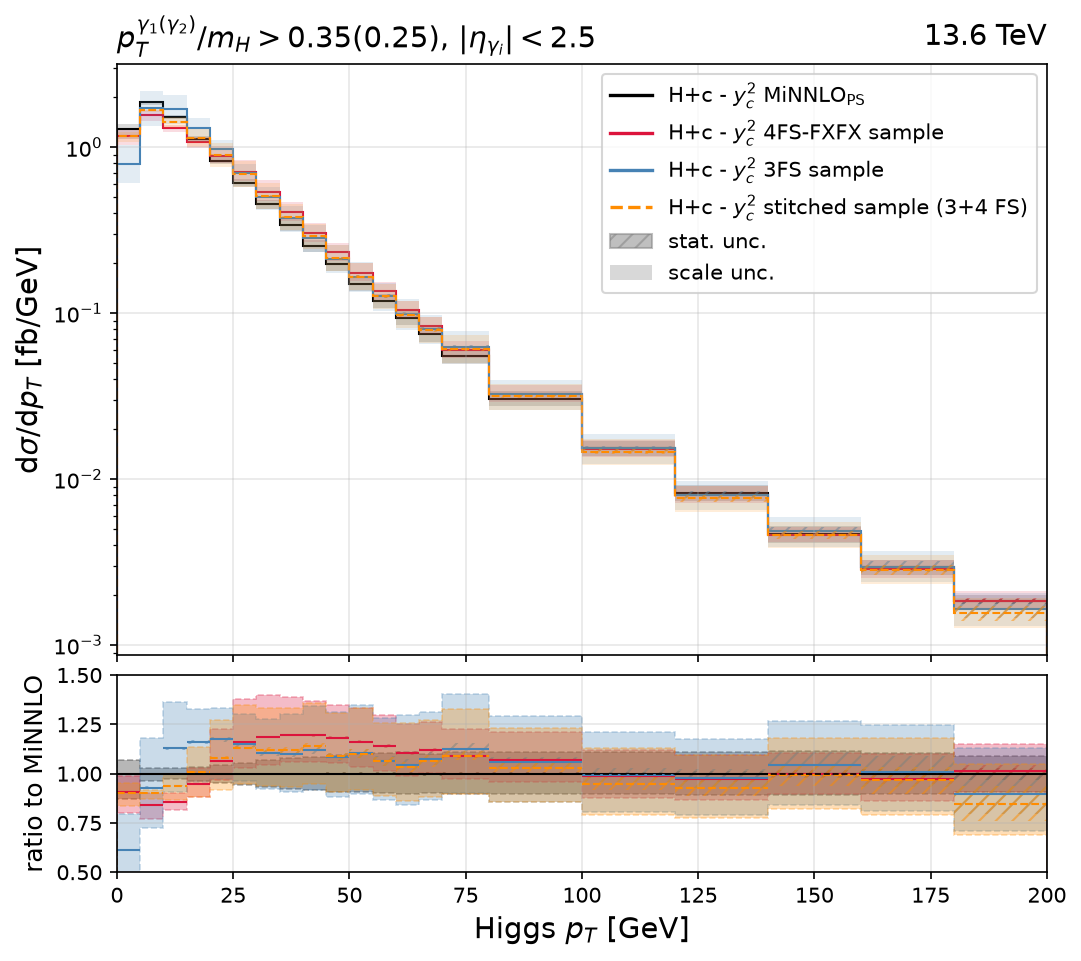}
    \includegraphics[width=0.49\textwidth]{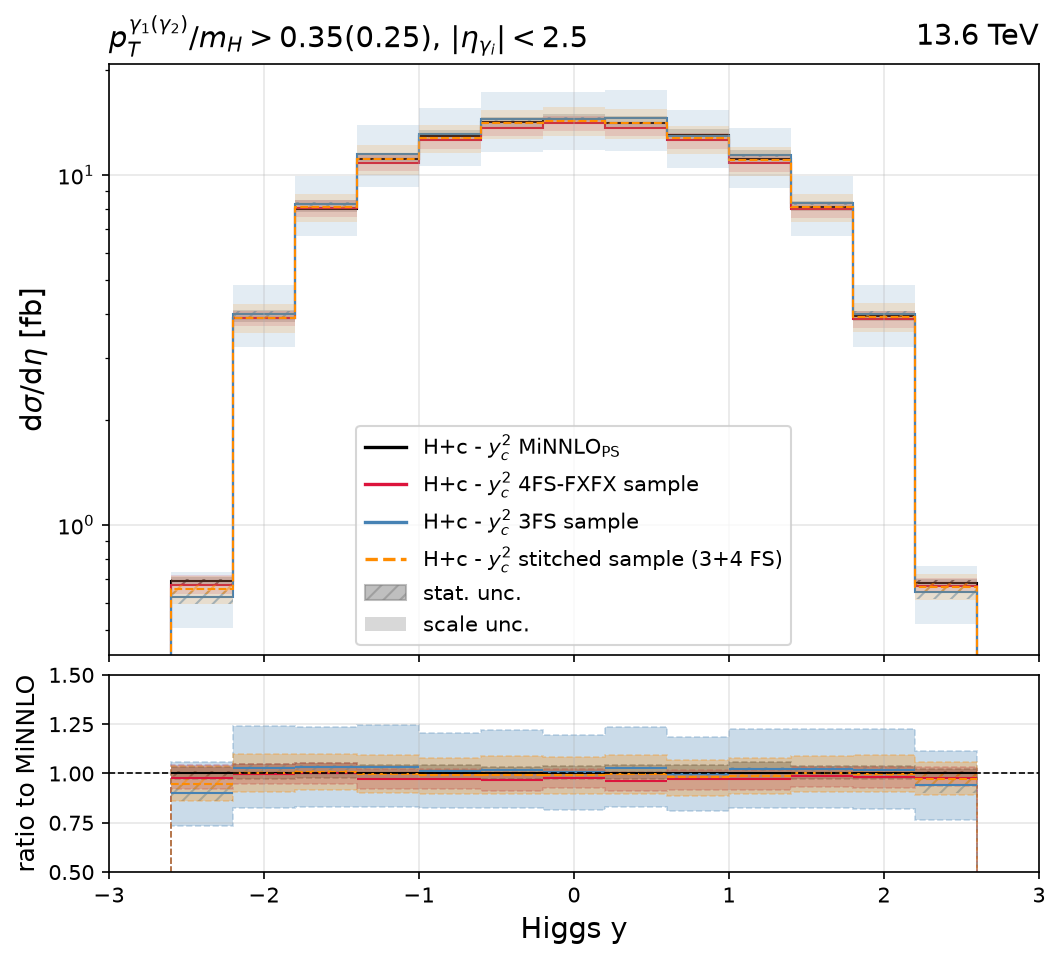}
    \includegraphics[width=0.49\textwidth]{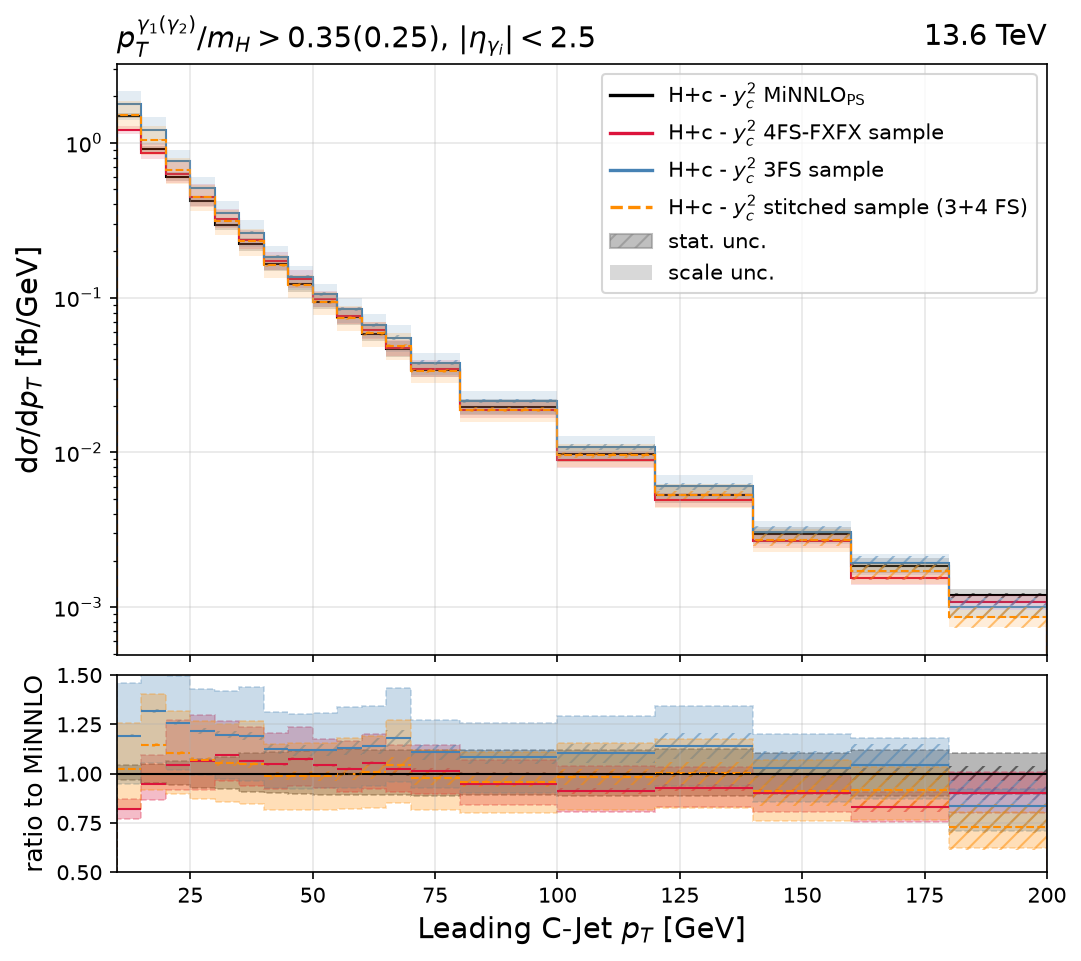}
    \caption{Differential $p_T(H)$, $y(H)$ and $p_T(\text{c-jet})$ cross section distribution comparison between the \minnlo{} sample and the 3FS, 4FS-FXFX and stitched samples. The three plot show the distributions requiring the presence of the leading and subleading photons from the Higgs Boson decay with $p^{\gamma_1,(\gamma_2)}_T/m_H > 0.35~(0.25)$ and $|\eta_\gamma|<2.5$, and a requirement on the $p_T$ of the c-jet to be greater than 10~GeV. For the \minnlo{} sample the fiducial volume selection efficiency is 63\% with respect to the inclusive selection.}
    \label{fig:MiNNLO_fid_aa}
\end{figure}

We now turn to differential results in fiducial phase spaces obtained by applying cuts on the photon decay products of the Higgs boson, with the decay simulated by \PYTHIA{}. The three panels of Figure~\ref{fig:MiNNLO_fid_aa} show distributions requiring two photons with $|\eta|<2.5$ and asymmetric transverse-momentum cuts, $p_T^{\gamma_1}/m_H > 0.35$ and $p_T^{\gamma_2}/m_H > 0.25$. We do not observe significant differences with respect to the inclusive case. In particular, both the Higgs-boson and leading-$c$-jet observables considered in this study exhibit a behaviour very similar to that observed in the inclusive analysis.\\

\begin{figure}[h!]
    \centering
    \includegraphics[width=0.49\textwidth]{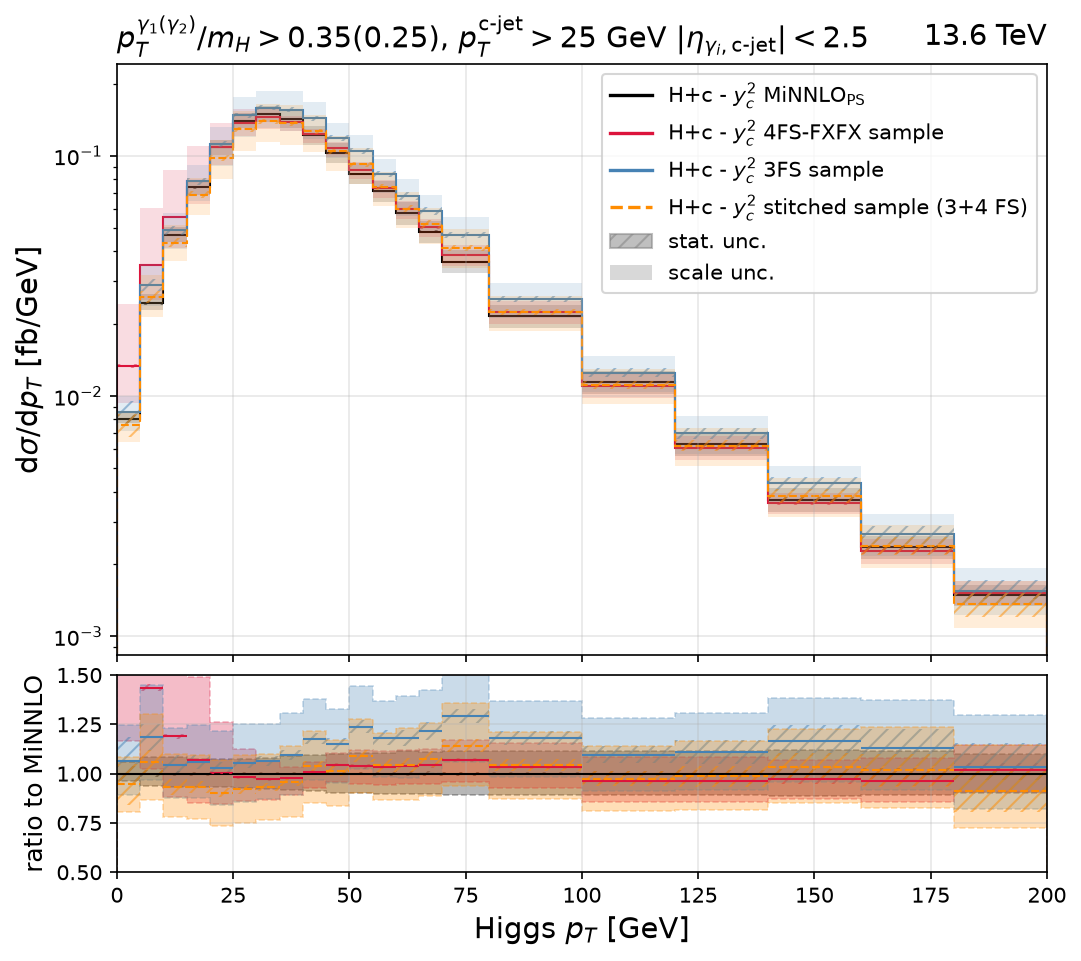} 
    \includegraphics[width=0.49\textwidth]{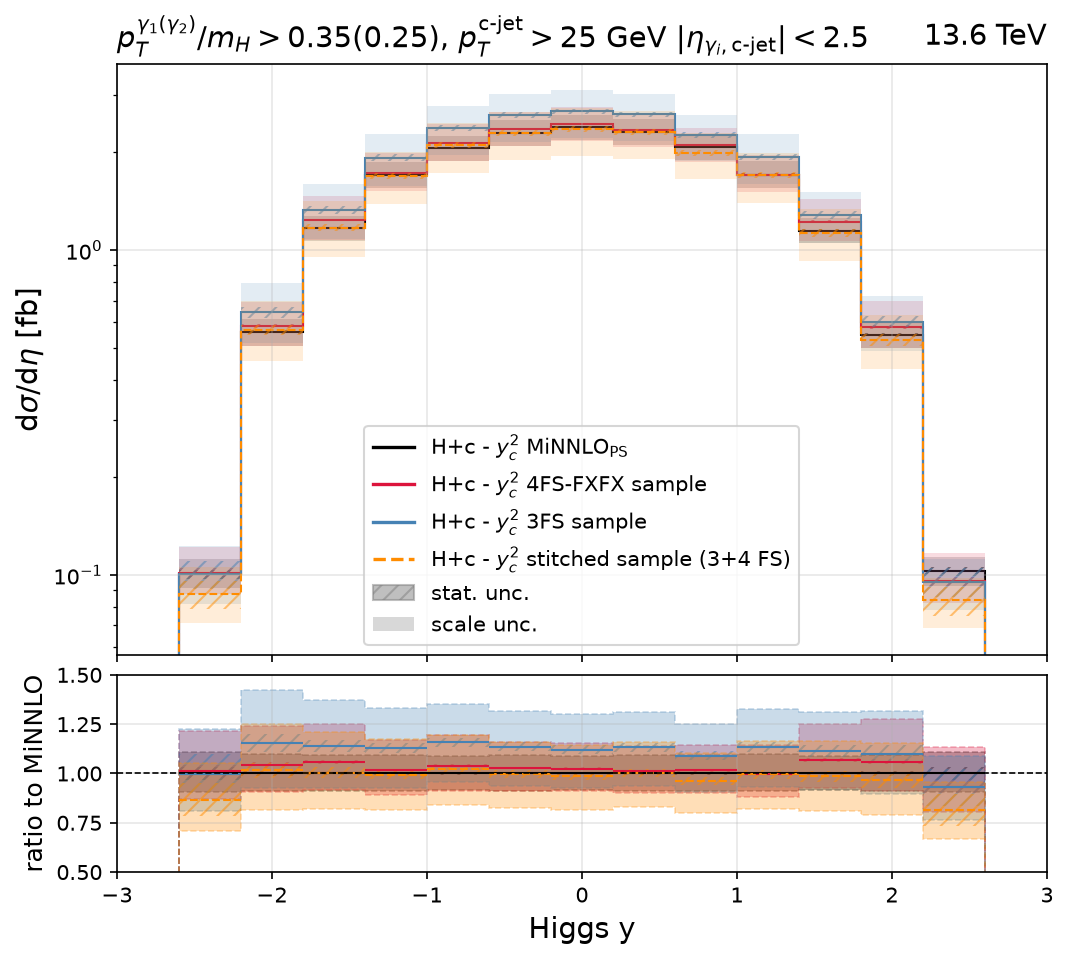}
    \includegraphics[width=0.49\textwidth]{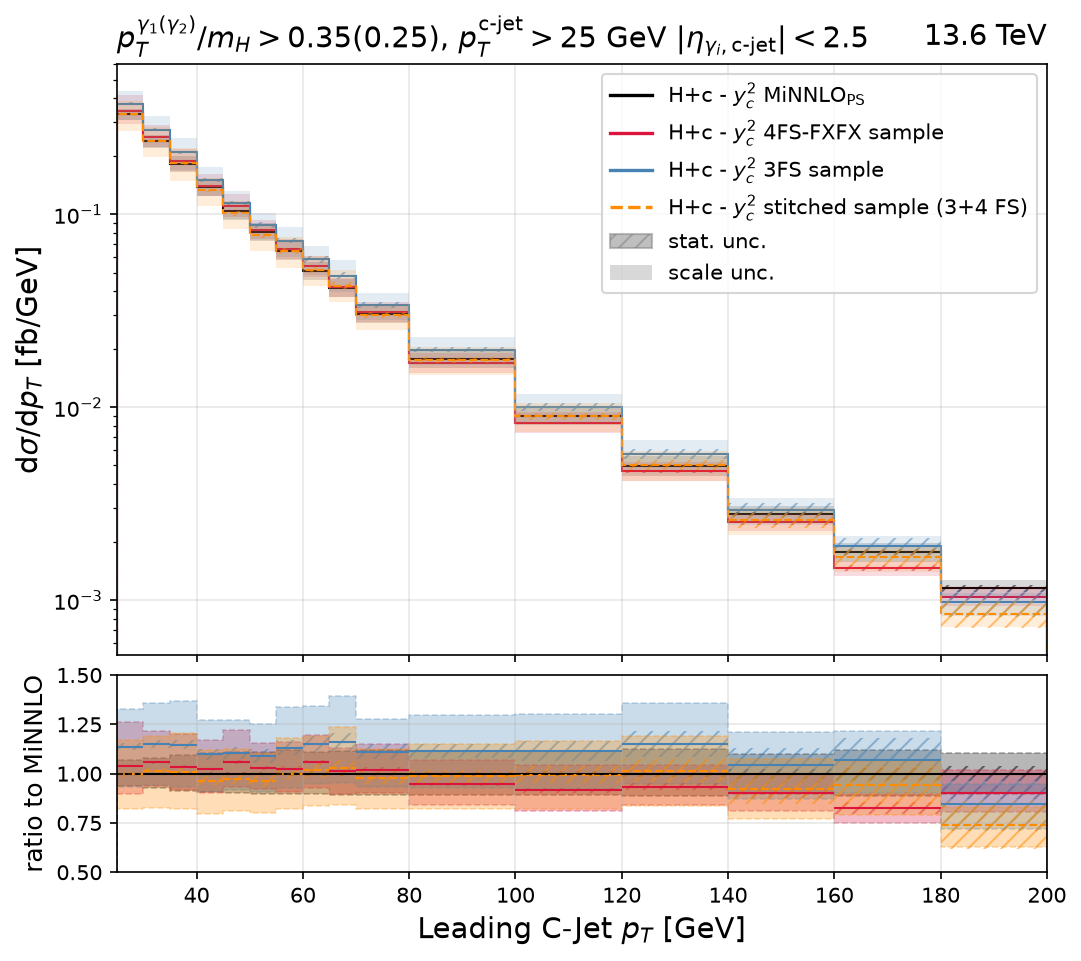}

    \caption{Differential $p_T(H)$, $y(H)$ and $p_T(\text{c-jet})$ cross section distribution comparison between the MiNNLOps sample and the 3FS, 4FS-FXFX and stitched samples. The three plot show the distributions requiring the presence of the leading and subleading photons from the Higgs Boson decay with $p^{\gamma_1,(\gamma_2)}_T/m_H > 0.35~(0.25)$ and $|\eta_\gamma|<2.5$, and the presence of a c-jet with $p_T>25$~GeV and $|\eta|<2.5$. For the \minnlo{} sample the fiducial volume selection efficiency is 10\% with respect to the inclusive selection.}
    \label{fig:MiNNLO_fid_aaj}
\end{figure}

We finally discuss the differential predictions in fiducial regions by imposing an additional selection cut for the leading $c$-jet. The three panels in Figure~\ref{fig:MiNNLO_fid_aaj} show the distributions requiring the presence of two photons with $|\eta| < 2.5$ and asymmetric $p_T$ cuts of $p_T^{{\gamma_1(\gamma_2)}}/m_H>0.35(0.25)$, and also requiring a c-jet with $p_T>25$~GeV and $|\eta| < 2.5$. For the Higgs-boson transverse-momentum distribution, we again observe good agreement between the \minnlo{} prediction and the massless FxFx merging in the high-$p_T$ tail, as expected. At low transverse momentum, however, the 4FS FxFx sample exhibits a different shape compared to both the massive NLO+PS and the massless \minnlo{} predictions, which are in better agreement with each other. In this region, the stitched sample closely follows the massive prediction and is therefore also closer to the \minnlo{} result, which incorporates higher-order and resummation effects. As a result, it provides a description that smoothly interpolates between the massive and massless predictions across the spectrum. On the other hand, for the rapidity distribution we observe good agreement between the \minnlo{} and 4FS-FxFx predictions. In this fiducial phase space, the massive prediction, normalised to the inclusive NNLO cross section, exhibits a nearly flat positive shift across the bulk of the rapidity spectrum, together with mild shape differences with respect to the massless predictions, particularly at large absolute rapidities. Nevertheless, these effects remain well within the scale-uncertainty bands of the predictions. Turning to the last panel of Figure~\ref{fig:MiNNLO_fid_aaj}, we observe a similar shape for all four predictions. For $p_T(H)$ grater than 20 GeV, the stitched result follows closely the 4FS-FxFx prediction, while remaining closer to the fiducial \minnlo{} result. As in the inclusive analysis, we observe a mild suppression of the 4FS-FxFx prediction with respect to the \minnlo{} simulation in the high-$p_T$ tail of the leading $c$-jet distribution.\\

Overall, the stitched sample improves both the kinematic description of the process and the scale uncertainty associated with the variation of the $\mu_R$ and $\mu_F$ starting values. The stitched distributions are closer to the NNLO and to the massive predictions in the low-$p_T$ regime, where charm-mass effects are most relevant, and compatible with both the NNLO and the massless predictions at high $p_T$, yielding an overall improvement in the ratio to the \minnlo{} sample with respect to both the 3FS and 4FS-FxFx distributions. The scale uncertainty bands are also smaller than for the non-stitched NLO samples, as the construction leverages the phase-space regions where each scheme performs best.\\

We recommend to perform NNLO+PS simulations  whenever \hpc production is a relevant process in the analysis, however in the absence of a full NNLO prediction at the analysis level, we suggest the use of the stitching procedure. Such approach mitigates the need for the flavour-scheme uncertainty obtained from a plain yield comparison of the massive and massless samples, as used in Ref.~\cite{CMS_chgg2025} and resulting in $\mathcal{O}(30\%)$ discrepancy after fiducial cuts, since the residual discrepancy with respect to the \minnlo{} shapes is covered by the scale uncertainty of the stitched sample. However, we stress the importance of reweighting the NLO+PS integrated cross-section with the most accurate inputs available, ideally from dedicated higher-order fixed-order calculations or simulations, such as the \minnlo{} predictions employed here.

%% file: sections/conclusions.tex
\section{Conclusions}

In this work, we have presented a comprehensive study of the simulation of Higgs-boson production in association with heavy-flavour quarks, with particular emphasis on charm-associated Higgs boson production. We have described the theoretical frameworks available for the simulation of $\PH+\PQc$ production, with particular emphasis on the different flavour schemes and their respective advantages and limitations. After discussing the interference and internal charm-loop contributions, we focused on the simulation of the dominant $y_c^2$ signal, where charm quarks are treated as asymptotic states in both the massless and massive flavour schemes. The NLO+PS simulations are performed within the \MADGRAPH framework, while the NNLO+PS simulations are obtained in \POWHEG{} using the \minnlo{} method. The massless NLO+PS predictions for $\PH$ and $\PH+j$ production are merged using the FxFx procedure. These results are then combined with the massive NLO+PS predictions through a stitched prescription that switches between flavour schemes according to the heavy-flavoured-jet multiplicity.

We have compared massive and massless flavour-scheme approaches, highlighting their respective domains of validity and quantifying their impact on inclusive and fiducial cross sections, as well as on differential observables. A detailed investigation of renormalisation, factorisation, and parton-shower scale choices has confirmed that the nominal setup adopted for \bbh production, based on a dynamic scale definition and moderate shower hardness, also provides good perturbative stability and consistency between schemes for \cch production. The different NLO+PS predictions, including the stitched sample, have been compared to the \minnlo{} 4FS calculation, both inclusively and in fiducial phase spaces defined by the Higgs-boson decay into photons and tagged $c$-jets. We find that the stitched prescription provides a good description of Higgs- and $c$-jet-related observables, particularly in the soft regions of phase space, where it improves upon the massless NLO+PS predictions.

A dedicated study of the charm-quark case has allowed us to provide first recommendations for the modelling of the $\PH+\PQc$ signal with NLO+PS generators, the treatment of the charm Yukawa coupling, and the interpretation of predictions obtained in the massive and massless flavour schemes. This work establishes a coherent and validated simulation framework for Higgs boson production in association with heavy-flavour quarks, complemented by practical prescriptions for estimating theoretical uncertainties, thereby supporting ongoing and future experimental efforts to probe the bottom- and charm-Yukawa couplings in Higgs-boson production at the LHC. Model files are available on request from the authors.

%% file: sections/Appendix.tex
\section{Scale dependence of the flavour-scheme tension}
\begin{figure}[ht!]
    \centering
    \includegraphics[width=0.495\textwidth, trim=0 0 0 65pt, clip]{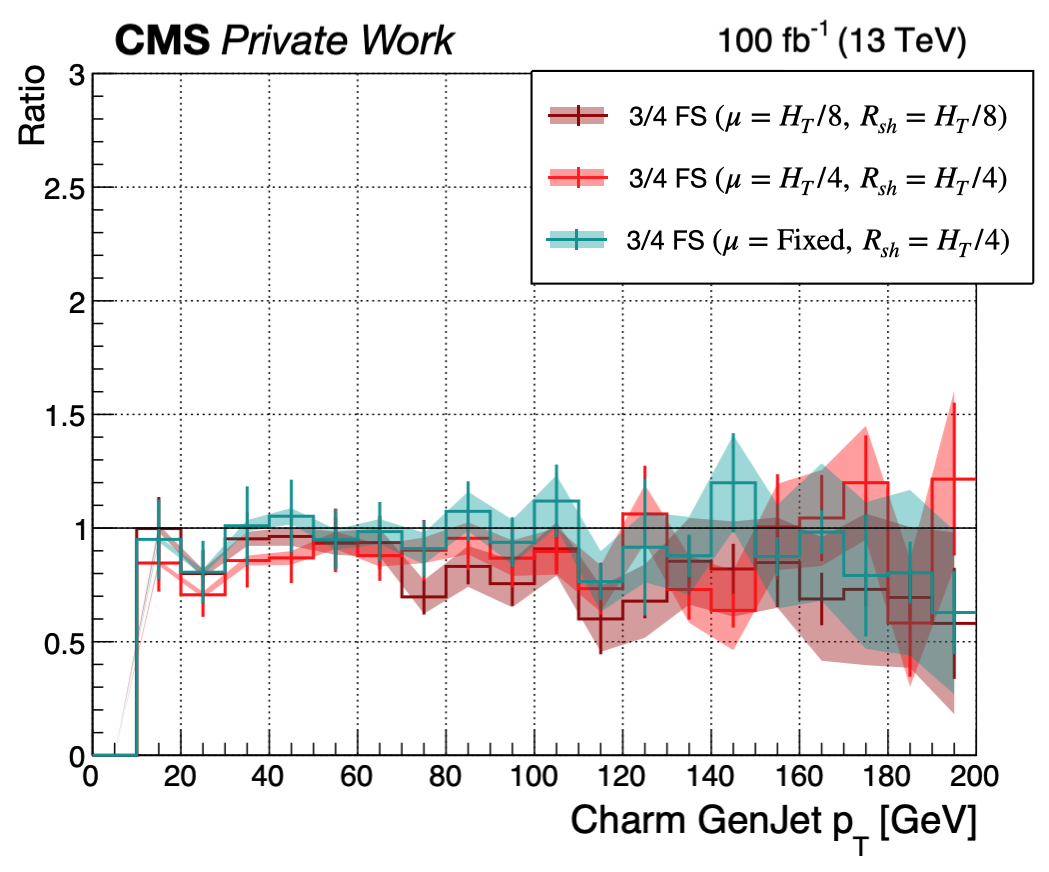}
    \includegraphics[width=0.495\textwidth, trim=0 0 0 65pt, clip]{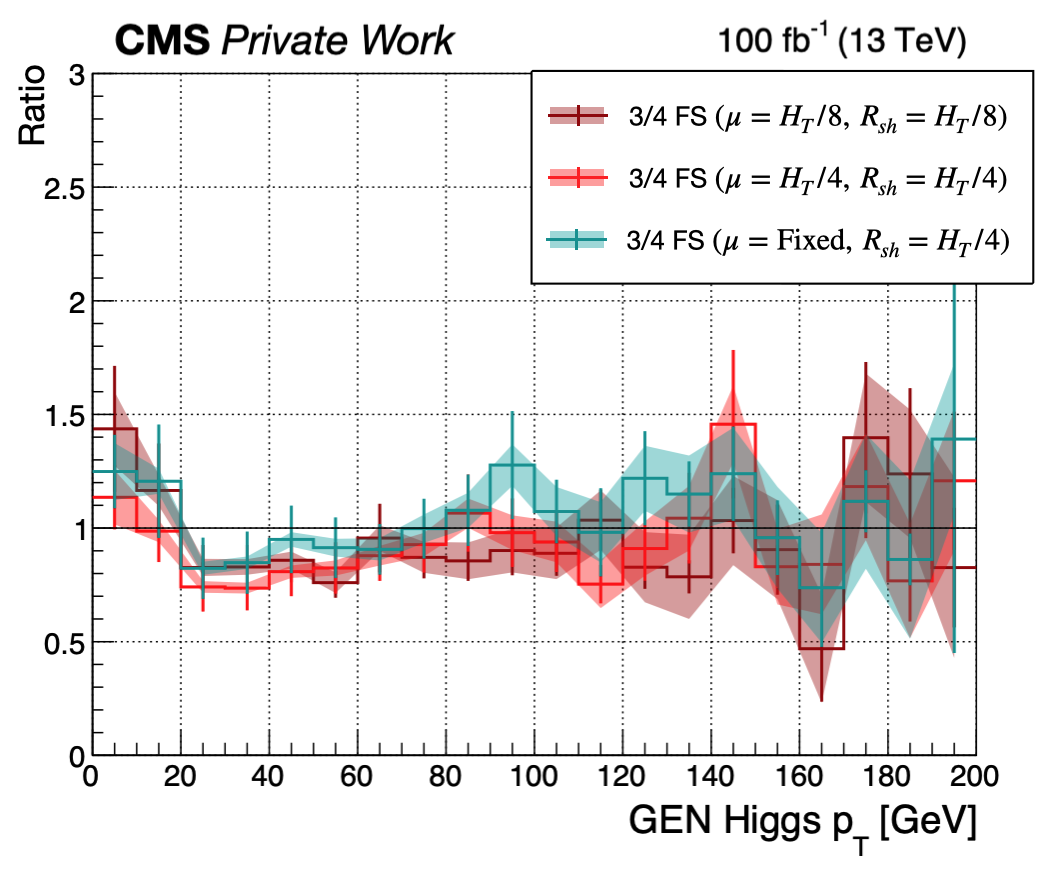}
    \caption{Ratio of 3FS to 4FS predictions for the $p_T$ of the leading GEN-$\PQc$-jet (left) and the Higgs boson (right).}
    \label{fig:ratios}
\end{figure}

In this appendix, we provide a study at 13.0 TeV on the scale dependence of the ratio between the cross-sections between the two factorisation schemes. Figure~\ref{fig:ratios} compares the ratio of 3FS to 4FS predictions for two observables: the $p_T$ of the hardest $\PQc$-jet (left) and the Higgs boson (right). For configurations yielding the smallest cross-section differences, we examined the flatness of these ratios, finding all three preferred configurations showing reasonable flatness. Based on this, the dynamic scale $\mu = H_T/4$ was chosen as the nominal setting for the signal sample, the fixed scale being disfavoured on theoretical grounds, and $\mu = H_T/8$ having larger scale uncertainties.\\

\begin{figure}[ht!]
    \centering
    \includegraphics[width=0.48\textwidth, trim=0 0 0 65pt, clip]{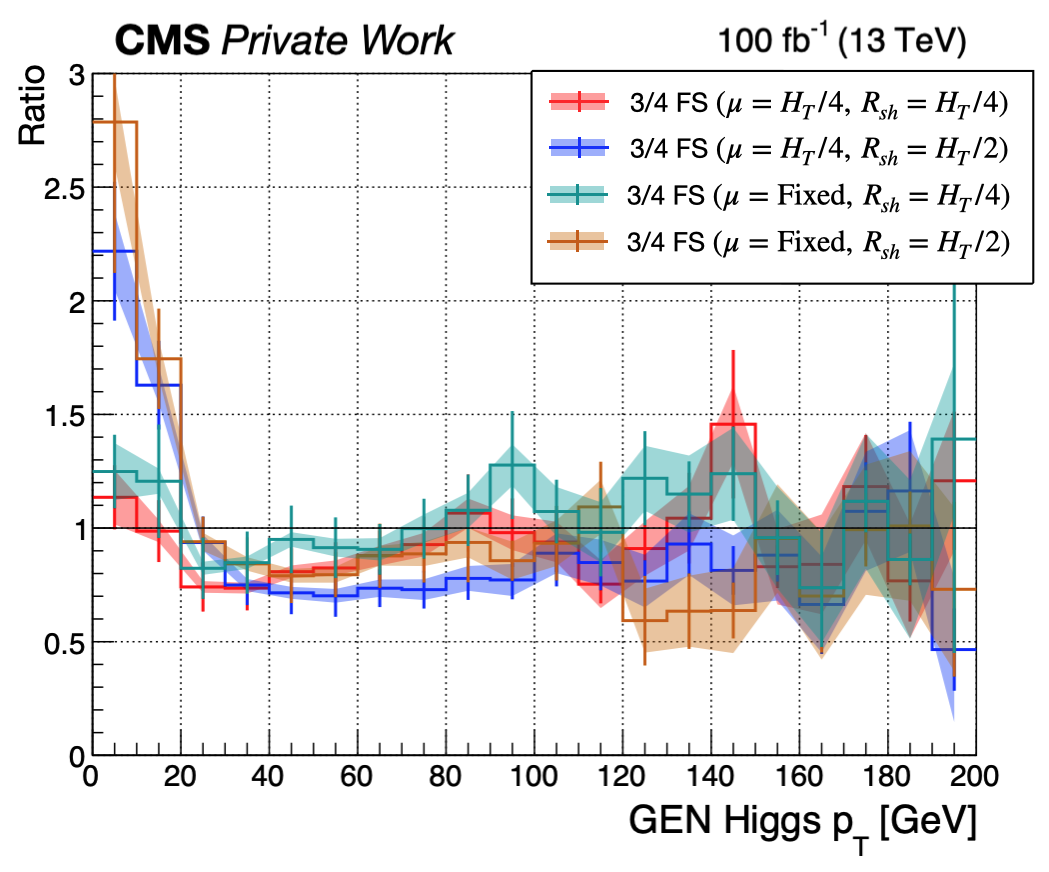}
    \caption{Impact of varying \texttt{shower\_scale\_factor}: 3FS/4FS ratio for Higgs $p_T$ distribution.}
    \label{fig:PSratios}
\end{figure}

Additionally, the optimal \texttt{shower\_scale\_factor} was determined by comparing differential distributions for samples with identical $\mu_R$ and $\mu_F$ values but varying PS scales. Figure~\ref{fig:PSratios} shows the impact on the Higgs boson $p_T$ distribution (left), and on the 3FS/4FS ratio (right). The setting \texttt{shower\_scale\_factor = 0.5} was selected as the nominal choice, as it minimizes the difference between 3FS and 4FS.

\section{Commands for NLO simulations and shower settings}\label{appendix}
In this appendix, we provide useful \MADGRAPH{} to reproduce the NLO+PS predictions presented in this work.

\subsection{$\PH+\PQc$ simulations} 
First, the \MADGRAPH{} commands for the simulation of $\PH+\PQc$ associated production across the various FS configurations explored in this study.\\

\noindent For the 3FS simulations:

\begin{verbatim}
    MG5_aMC> import model loop_sm_MSbar_yb_yc-yc3FS
    MG5_aMC> define p = g u u~ d d~ s s~
    MG5_aMC> define j = g u u~ d d~ s s~
    MG5_aMC> generate p p > h c c~ [QCD]
\end{verbatim}
    
\noindent For the 4FS simulations:
    
\begin{verbatim}
    MG5_aMC> import model loop_sm_MSbar_yb_yc-yc4FS
    MG5_aMC> define p = g u u~ d d~ s s~ c c~
    MG5_aMC> define j = g u u~ d d~ s s~ c c~
    MG5_aMC> generate p p > h [QCD]
\end{verbatim}

\noindent For the 4FS with FxFx merging:

\begin{verbatim}
    MG5_aMC> import model loop_sm_MSbar_yb_yc-yc4FS
    MG5_aMC> define p = g u u~ d d~ s s~ c c~
    MG5_aMC> define j = g u u~ d d~ s s~ c c~
    MG5_aMC> generate p p > h [QCD] @0
    MG5_aMC> generate p p > h j [QCD] @1
\end{verbatim}

\noindent These are followed by the standard output and launch commands.

\subsection{$\PH+\PQb$ simulations}
Similarly, we report the commands for the NLO+PS simulation of $\PH+\PQb$ associated production in the various FS configuration in \MADGRAPH{}.\\

\noindent For the 4FS:

\begin{verbatim}
    MG5_aMC> import model loop_sm_MSbar_yb_yc-yb4FS
    MG5_aMC> define p = g u u~ d d~ s s~ c c~
    MG5_aMC> define j = g u u~ d d~ s s~ c c~
    MG5_aMC> generate p p > h b b~ [QCD]
\end{verbatim}
    
\noindent For the 5FS:
    
\begin{verbatim}
    MG5_aMC> import model loop_sm_MSbar_yb_yc-yb5FS
    MG5_aMC> define p = g u u~ d d~ s s~ c c~ b b~
    MG5_aMC> define j = g u u~ d d~ s s~ c c~ b b~
    MG5_aMC> generate p p > h [QCD]
\end{verbatim}

\noindent For the 5FS with FxFx merging:

\begin{verbatim}
    MG5_aMC> import model loop_sm_MSbar_yb_yc-yb5FS
    MG5_aMC> define p = g u u~ d d~ s s~ c c~ b b~
    MG5_aMC> define j = g u u~ d d~ s s~ c c~ b b~
    MG5_aMC> generate p p > h [QCD] @0
    MG5_aMC> generate p p > h j [QCD] @1
\end{verbatim}

\noindent These are followed by the standard output and launch commands.

\subsection{Higgs emission from charm-quark loops}

We now report the commands for the simulation of charm-quark loop induced production in the 3FS configuration explored in this study:

\begin{verbatim}
    MG5_aMC> import model loop_sm_MSbar_yb_yc-yc3FS
    MG5_aMC> define lj = g u u~ d d~ s s~
    MG5_aMC> generate g g > h [noborn=QCD]
    MG5_aMC> add process lj lj > h lj [noborn=QCD]
\end{verbatim}

\subsection{Parton-shower settings}

As anticipated, the modelling of parton showering, hadronization, and underlying-event effects is performed with \PYTHIA{}8, using version 8.240 for the NLO samples and version 8.244 for the \minnlo{} samples. The specific settings employed in all matched simulations are reported below:
\begin{itemize}
    \item \texttt{JetMatching:scheme = 1}
    \item \texttt{JetMatching:qCut = 30.0}
    \item \texttt{JetMatching:qCutME = 10.0}
    \item \texttt{JetMatching:nQmatch = 4}
    \item \texttt{JetMatching:nJetMax = 1}
    \item \texttt{Tune:pp 14},
    \item \texttt{MultipartonInteractions:ecmPow=0.03344}
    \item \texttt{MultipartonInteractions:bProfile=2}
    \item \texttt{MultipartonInteractions:pT0Ref=1.41}
    \item \texttt{MultipartonInteractions:coreRadius=0.7634}
    \item \texttt{MultipartonInteractions:coreFraction=0.63}
    \item \texttt{MultipartonInteractions(TimeShower):alphaSvalue=0.118}
    \item \texttt{MultipartonInteractions(TimeShower):alphaSorder=2}
    \item \texttt{SigmaTotal:mode = 0}   
    \item \texttt{SigmaTotal:sigmaEl = 22.08}    
    \item \texttt{SigmaTotal:sigmaTot = 101.037}
    \item \texttt{Space(Time)Shower:pTmaxMatch = 1}
    \item \texttt{Space(Time)Shower:pTmaxFudge = 1}
    \item \texttt{Space(Time)Shower:MEcorrections = off}
    \item \texttt{TimeShower:globalRecoil = on}
    \item \texttt{TimeShower:limitPTmaxGlobal = on}
    \item \texttt{TimeShower:nMaxGlobalRecoil = 1}
    \item \texttt{TimeShower:globalRecoilMode = 2}
    \item \texttt{TimeShower:nMaxGlobalBranch = 1}
\end{itemize}